\documentclass[a4paper,preprintnumbers,nofootinbib,twocolumn]{quantumarticle}
\pdfoutput=1
\usepackage[english]{babel}
\usepackage{amssymb}
\usepackage{amsfonts}
\usepackage{amsmath}
\usepackage{eucal}
\usepackage{graphicx,color}
\usepackage{epsfig}
\usepackage{tikz}
\usepackage{cancel}
\usepackage{hyperref}
\usepackage[absolute,overlay]{textpos}

\usetikzlibrary{ arrows, babel, calc,decorations.pathmorphing, decorations.markings, decorations.pathreplacing, fit, positioning, shapes }

\newcommand{\be}{\begin{equation}} \newcommand{\ee}{\end{equation}}
\newcommand{\bea}{\begin{eqnarray}} \newcommand{\eea}{\end{eqnarray}}
\newcommand{\beann}{\begin{eqnarray*}}  \newcommand{\eeann}{\end{eqnarray*}}
\newcommand{\bfig}{\begin{figure}} \newcommand{\efig}{\end{figure}}
\newcommand{\ba}{\begin{array}} \newcommand{\ea}{\end{array}}
\newcommand{\bcen}{\begin{center}} \newcommand{\ecen}{\end{center}}
\newcommand{\btab}{\begin{tabular}} \newcommand{\etab}{\end{tabular}}

\usepackage{amsmath,mathtools,dsfont}
\newcommand{\id}{\mathds 1}

\def\barray{\begin{eqnarray}}
\def\earray{\end{eqnarray}}
\def\beq{\begin{equation}}
\def\eeq{\end{equation}}

\newtheorem{Proposition}{Proposition}[section]

\newtheorem{Theorem}{Theorem}[section]
\newtheorem{Lemma}{Lemma}[section]
\newtheorem{Corrolary}{Corrolary}[section]

\newcommand{\bp}{\begin{Proposition}}	\newcommand{\ep}{\end{Proposition}}
\newcommand{\bt}{\begin{Theorem}}	\newcommand{\et}{\end{Theorem}}
\newcommand{\bl}{\begin{Lemma}}		\newcommand{\el}{\end{Lemma}}
\newcommand{\bc}{\begin{Corrolary}}	\newcommand{\ec}{\end{Corrolary}}

\begin{document}

\title{Tensor Renormalization Group for interacting quantum fields}

\author{Manuel Campos, Germ\'an Sierra and Esperanza L\'opez}
\affiliation{
Instituto de F\'{\i}sica Te\'orica UAM/CSIC, C/ Nicol\'as Cabrera 13-15, Cantoblanco, 28049 Madrid, Spain}

\begin{abstract}
We present a new tensor network algorithm for calculating the partition function of interacting quantum field theories in 2 dimensions. It is based on the Tensor Renormalization Group (TRG) protocol, adapted to operate entirely at the level of fields. This strategy was applied in Ref.[1] to the much simpler case of a free boson, obtaining an excellent performance. Here we include an arbitrary self-interaction and treat it in the context of perturbation theory. A real space analogue of the Wilsonian effective action and its expansion in Feynman graphs is proposed. Using a $\lambda \phi^4$ theory for benchmark, we evaluate the order $\lambda$ correction to the free energy. The results show a fast convergence with the bond dimension, implying that our algorithm captures well the effect of interaction on entanglement.

\end{abstract}

\begin{textblock*}{5cm}(15.1cm,1.2cm) 
   IFT-UAM/CSIC-21-46
\end{textblock*}
\maketitle


Tensor network techniques have been crucial to study strongly coupled spin systems relevant to condensed matter
and statistical physics \cite{A88}$-$\cite{V21}. The application of these techniques to quantum fields presents the challenge of dealing with
infinite dimensional systems. Several strategies have been pursued to this aim. The most obvious is to introduce a truncation
that brings back to the simpler framework of finite dimensional Hilbert spaces. An opposite philosophy
has lead to the development of continuous versions of tensor networks 
\cite{V10}$-$\cite{T21b}.
Tensor networks have also provided simple versions of the holographic principle within the AdS/CFT
correspondence \cite{Sw12}$-$\cite{J21}. 

Inspired on standard field theory techniques, a different strategy was explored in \cite{CS19}. The main idea 
was to keep the continuous character of fields, treating them as the basic element for the implementation of an adapted
Tensor Renormalization Group (TRG) protocol \cite{LN07}. The viability of this approach was tested by evaluating the partition function 
of a free massive boson on a 2-dimensional square lattice. The results exhibited a very good numerical precision with moderate 
bond dimension, which in this case counts the number of fields per lattice link. Remarkably the massless limit could be addressed 
without an increase in the bond dimension.
 
The strength of the method relies on the combination of semi-analytic expressions and numerics that only involves 
finite dimensional matrices. This however is heavily based on the gaussian nature of free 
fields. Therefore it was an open question whether the same ideas 
are applicable to interacting field theories. We show 
that the answer is affirmative. Moreover, the semi-analytical character of the method 
exhibits 
important aspects of the interplay 
between interaction and entanglement.

The paper is organized as follows. We set the basis of our proposal in Section I. Section II reviews the application to free fields. Section III contains a brief summary of the Wilsonian approach
to quantum fields theories.
Interaction is introduced in Section IV in the context of perturbation theory. The structure of the TRG for interacting quantum fields is analized in Section V. 
We explore the numerical performance of the protocol in Section VI.
Section VII presents a summary of results and discusses further lines of research.
Technical details are consigned to appendices.
An implementation in Mathematica of this adapted TRG newtwork can be found in the GitHub repository \cite{repG}. 

\section{A Tensor network for quantum fields}

The aim of this paper is to present a new framework for a real space renormalization group (RG) analysis of quantum field theories.
It will combine the main guidelines of a conventional particle physics approach to quantum field theory \cite{W75,S94} with the introduction
of a classification of degrees of freedom based on entanglement. 
The main focus of study will be the partition function, whose
RG analysis will be formulated in terms of fields $x \in \mathbb{R}$. In the spirit of lattice field theory, the spacetime
will be discretized.

We will work in two dimensions and consider a scalar field theory with Lagrangian
\be
{\cal L}= {1 \over 2} (\partial x)^2 + {1 \over 2} m_0^2 x^2 + \lambda_0 {\rm v} (x) \ ,
\label{lagrangian}
\ee
where $m_0$ is the boson mass and $\lambda_0$ a coupling constant.
For concrete numerical evaluations we will choose ${\rm v}(x)=x^4$, but otherwise we keep the potential generic.
We will study a discretized version of this theory living on a square lattice.
The field variables are assigned to the lattice links while the statistical or Boltzmann weights are carried by the vertices
\vspace*{-3.5cm}
\begin{equation}
\hspace{0mm}  \vcenter{\hbox{\includegraphics[width=5.7cm]{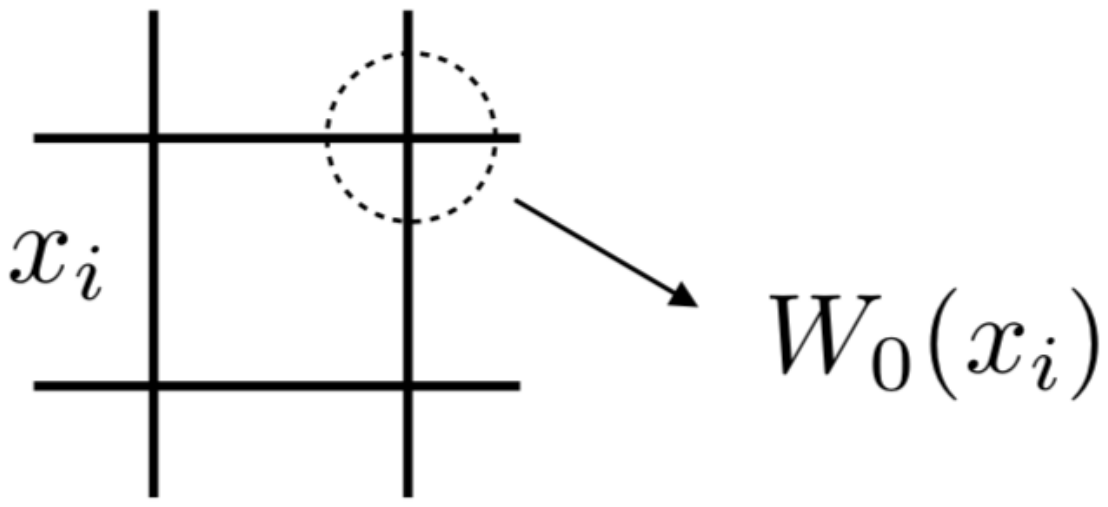}} }  \hspace{-10mm} \ .
\end{equation}
\vspace*{-3.6cm}

\noindent The weights $W_0(x_i)$ are given by
\vspace*{-3.6cm}
\be
\hspace{-30mm}  \vcenter{\hbox{\includegraphics[width=5.8cm]{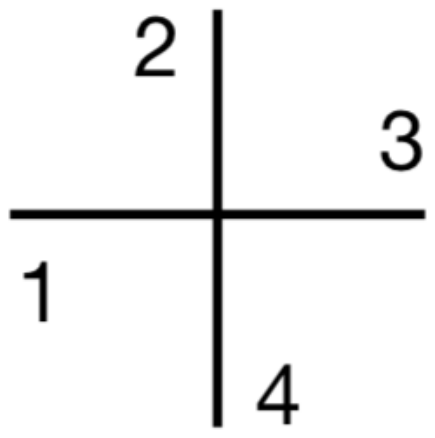}} } \hspace{-23mm}
\, = \, e^{-{1 \over 2} \sum_{i=1}^4 \big[ (x_i -x_{i+1})^2 + {m^2 \over 2} x_i^2 +  \lambda {\rm v}(x_i) \big]}  \ ,
\label{Wint}
\ee
\vspace*{-3.9cm}

\noindent 
where we are considering a unit lattice spacing, hence $m=m_0$ and $\lambda=\lambda_0$.
The partition function of the so defined vertex model is
\be
Z=\int \prod_{i \in {\rm links}}  d x_i \;\, \times \!\!\!\!\! \prod_{j \in {\rm vertices}} \!\!\! W_{0j}(x_i) \ .
\label{Z}
\ee

The evaluation of the partition function will be done by means of an adapted Tensor Renormalization Group protocol (TRG) that reduces iteratively the size of the lattice. 
The TRG  for finite dimensional systems is based on the singular value decomposition (SVD) of the Boltzmann weights \cite{LN07}
\be
W=U_L\, D\, U_R^+ \ , 
\ee
where $U_{L,R}$ are unitary matrices, with $U_R^+$ denoting the adjoint of $U_R$, and $D$ is a diagonal matrix containing the singular values of $W$.
This decomposition has the following graphical representation
\vspace*{-1.8cm}
\begin{equation}
  \vcenter{\hbox{\includegraphics[width=7.7cm]{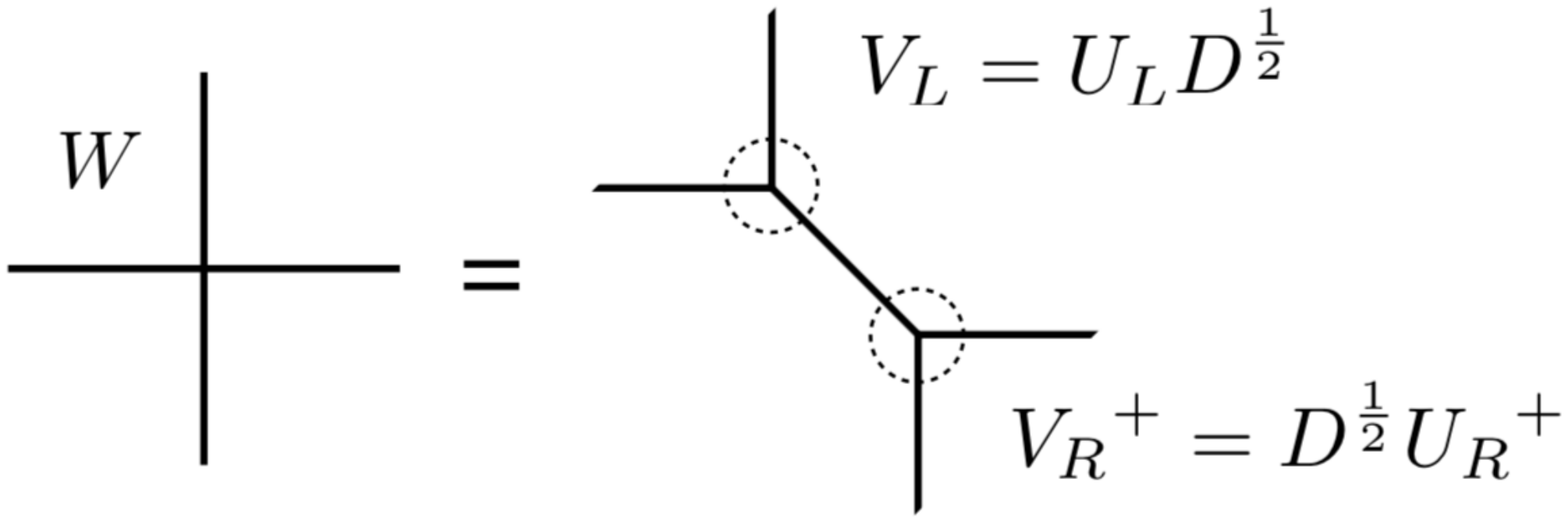}} }   \hspace{-10mm} \ .
\label{TRGfig}
\end{equation}
\vspace*{-2.1cm}

\noindent
Namely, the four-leg Boltzmann weight $W$ is decomposed in two cubic weights $V$. This decomposition can be analogously performed along the opposite diagonal.
The variables in the new tilted links have a one to one correspondence with the singular values of $W$. Therefore they
have a natural hierarchy based on entanglement. This allows to discard those with the weakest 
contribution. After the splitting \eqref{TRGfig} the old variables are summed over, eliminating degrees of freedom associated with short range entanglement.
The number of lattice links is reduced in this process by a factor of two
\vspace*{-3.6cm}
\begin{equation}
  \vcenter{\hbox{\includegraphics[width=6cm]{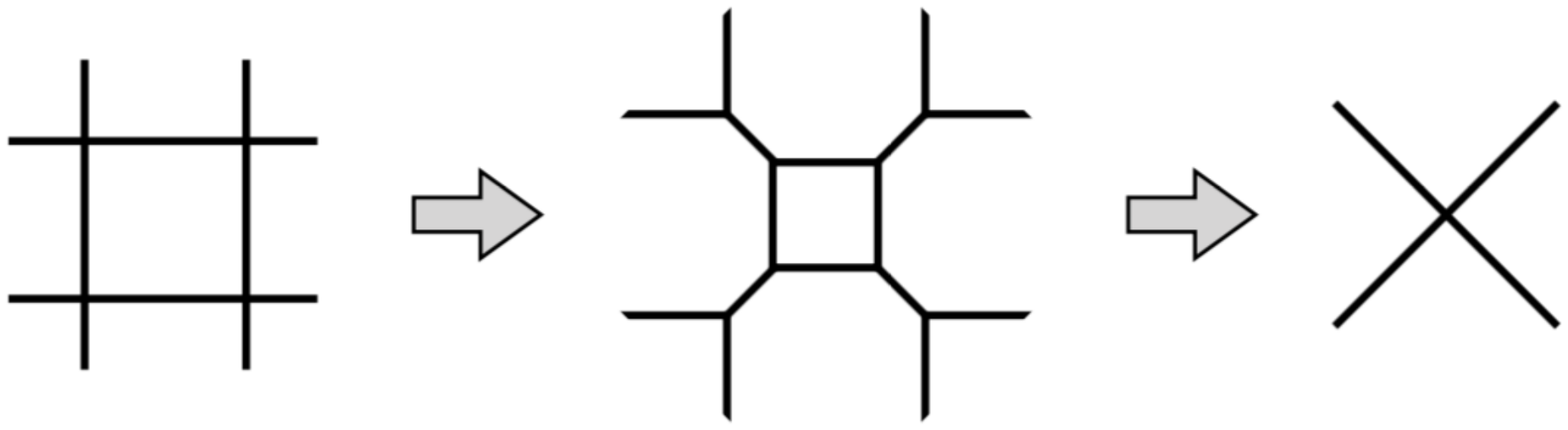}}}  \hspace{0mm} \ .
\label{TRGfig2}
\end{equation}
\vspace*{-3.6cm}

Our proposal of a tensor network for quantum field theories can be summarized in a simple idea: apply the SVD splitting not to the Boltzmann weights,
but to their exponent. 
This allows to treat fields as the basic elements, and to deal with finite tensors instead of functions while preserving the continuous character of the Boltzmann weights.

\section{TRG for free fields}

Let $W_n$ denote the Boltzmann weights of the discretized boson model at a given coarse graining level
\vspace*{-3.5cm}
\begin{equation}
\hspace{-30mm}  \vcenter{\hbox{\includegraphics[width=6cm]{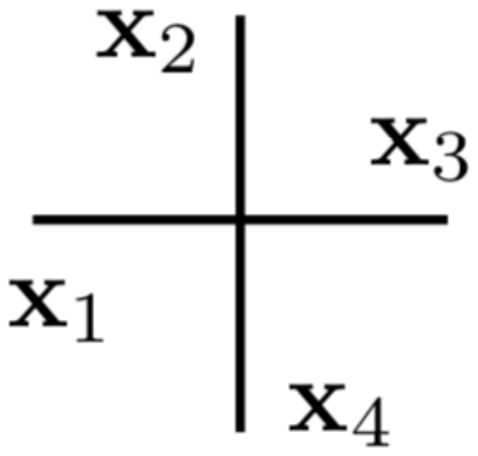}}} \hspace{-20mm} = \;\; W_n({\bf x}_i) \ .
\label{WW}
\end{equation}
\vspace*{-3.8cm}

\noindent where ${\bf x}_i=(x_{i1},\dots, x_{i \chi_n})$ is the set of fields carried by each lattice link. We will
refer to $\chi_n$ as the bond dimension of the tensor network.
In analogy with \eqref{TRGfig}, we are searching for a factorization of the form
\be
W_n({\bf x}_i) = \int d {\bf p} \; V_{n}({\bf x}_1 ,{\bf x}_2;{\bf p}) \, V_{n}^+ ({\bf p};{\bf x}_4,{\bf x}_3) \ .
\label{factLR}
\ee
The variables ${\bf p}=(p_{1},\dots, p_{\chi_{n+1}})$, necessary for the splitting of $W_n$, are associated with the internal tilted link in  \eqref{TRGfig}.
We require that $p_j \in \mathbb{R}$ in order to interpret them as fields. Thus $\chi_{n+1}$ will be the bond dimension of the new links.
In this simple implementation of the TRG, $x$-fields on straight links and $p$-fields on tilted links alternate along the coarse graining iterations.
The symmetries of the model allow to make a choice of basis for the fields such that all cubic 
lattice
weights, independently of their orientation, share the same expression.
This choice has been assumed in \eqref{factLR}.

\subsection{Gaussian SVD}

We consider first the implementation of this approach on 
the initial weights $W_0$ \eqref{Wint}. 
The mass and interaction terms in the Lagrangian \eqref{lagrangian} do not contribute to the coupling of fields across lattice links. 
The only obstruction to a trivial splitting of $W_0$ is the factor
\be
e^{-{1 \over 2} \big[ (x_2-x_3)^2 + (x_1-x_4)^2 \big]} \ .
\label{mix0}
\ee
Remarkably a simple Fourier transform is enough to realise the scheme proposed in \eqref{factLR}
\be
e^{-{1 \over 2} (x_1-x_4)^2 } = {1 \over \sqrt{2 \pi}} \int d p_1 \, e^{\, i p_1 (x_1 -x_4) -{1 \over 2} p_1^2} \ .
\label{fourier}
\ee
An analogous expression holds for the mixing between $x_{2,3}$ after introducing a second splitting field $p_2$. Hence the bond dimension is doubled in this process. 
While a single field lives at the links of the original lattice, $\chi_0=1$, the new tilted links carry  two, $\chi_1=2$. 

This simple mechanism is enough for the real space RG analysis of free field theories \cite{CS19}. In the absence of interaction, the weights \eqref{Wint} are
gaussian. The splitting \eqref{fourier} respects this property. We can thus write the Boltzmann weights at any coarse graining level $n$ as
\be
W_n({\bf x})= F_n \, e^{-{1 \over 2} {\bf x} \, M_n \, {\bf x}} \ ,
\label{Wgenfree}
\ee
with $M_n$ a $4 \chi_n \times 4 \chi_n$ matrix and $F_n$ a numerical factor.
From now on boldface symbols without indices will denote collectively all fields involved in the expression under consideration.
The argument of the Boltzmann weights are $x$-type fields for $n$ even and  $p$-type fields for $n$ odd. For convenience and without loss of generality we have chosen the former 
to write down the general expression \eqref{Wgenfree}.  
A basis can be chosen  \cite{CS19} where $M_n$ has the form 
\be
M_n=\begin{pmatrix} A_n & 0 \\
                              0 & A_n \end{pmatrix} +  \begin{pmatrix}\;\;\;B_n  & -B_n \\
                                   \!\!  -B_n & \;\;\;B_n \end{pmatrix} \ ,                                    
\label{Mfree}                                  
\ee                   
where $A_n$ and $B_n$ are real, non-negative, symmetric matrices, and $M_n$ acts on ${\bf x}=({\bf x}_1,{\bf x}_2,{\bf x}_4,{\bf x}_3)$. 
For the initial weights $W_0$ the previous matrices are given by
\be
A_0=\begin{pmatrix} \;\;\;1& \!\!\!-1 \\
                             \! -1 &\;\; 1 \end{pmatrix} + \;{m^2 \over 2} \, \id_2
                               \ , \hspace{5mm}  B_0= \id_2 \ .
\label{A0B0}                               
\ee

The $2 \times 2 $ block structure of \eqref{Mfree} reflects the LR factorization that we need to implement in order to move to the next coarse graining level.
With respect to the splitting \eqref{TRGfig}, we have ${\bf x}=({\bf x}_L,{\bf x}_R)$ with ${\bf x}_L=({\bf x}_1,{\bf x}_2)$ and ${\bf x}_R=({\bf x}_4,{\bf x}_3)$.
Generalizing \eqref{mix0}, the piece of the Boltzmann weights responsible for LR mixing is
\be
e^{-{1 \over 2} ({\bf x}_L-{\bf x}_R)\, B_n ({\bf x}_L-{\bf x}_R)} \ .
\label{B}
\ee
Because of the mentioned properties of $B_n$,
its SVD reduces to $B_n=U_n D_n U_n^T$ with $U_n$ an orthogonal matrix.
Using this decomposition, we can apply again a Fourier transformation  to  rewrite  \eqref{B} as
\be
{1 \over \sqrt{(2\pi)^{\chi_{n+1}} \det D_n}} \int d {\bf p} \, e^{\,i ({\bf x}_L-{\bf x}_R)\, U_n \,{\bf p}\, -{1 \over 2} {\bf p} \, D_n^{-1} {\bf p}} \ ,
\label{fourier2}
\ee
which implements the desired factorization. 
The splitting fields ${\bf p}$ are in one to one correspondence with the singular values of $B_n$.
In this way the TRG machinery is moved from the Boltzmann weights to their exponent, where fields are treated as the basic elements. 

We add for completeness the expression of $V_n$
\begin{equation}
\vcenter{\hbox{\includegraphics[trim=8cm 12.5cm 7.5cm 12.5cm, clip, width=1.5cm]{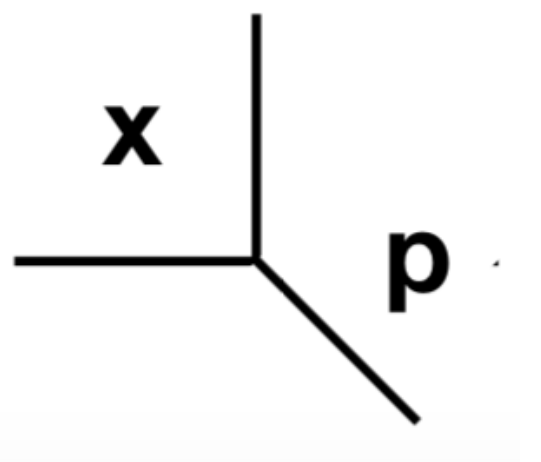}}} 
=\;\;\; f_n \; e^{-{1\over 2} {\bf x} \, A_n \,{\bf x}\, +\,i {\bf x}\, U_n {\bf p}\, -{1\over 4} {\bf p} \, D_n^{-1} {\bf p}} \ .
\label{cubicfree}
\end{equation}
\noindent It holds for cubic weights with any of the four possible orientations.
The normalization constant $f_n$ is defined by the relation
\be
F_n= \rho_n \, f_n^2 \ , \hspace{10mm}  \rho_n=   \sqrt{(2\pi)^{\chi_{n+1}} \det D_n} \ .
\label{fr}
\ee

This factorization mechanism was named gaussian SVD \cite{CS19}. In Appendix A we analyze the relation between the gaussian and the 
standard SVD, which for the Boltzmann weights \eqref{Wgenfree} can be carried out explicitly. We show that the latter is also governed by the singular values
of the matrix $B_n$. Therefore the gaussian SVD is not a different way of evaluating the LR mixing, which the standard SVD has been proved to do in an optimal way. 
Instead it provides a method to organize the infinite set of singular values obtained when working with continuous functions, grouping them into a finite number of subsets associated with fields. 

We include a summary of the first two steps of the TRG 

\vspace*{-1.2cm}
\bea
&& \hspace{-0mm}  \vcenter{\hbox{\includegraphics[width=8.2cm]{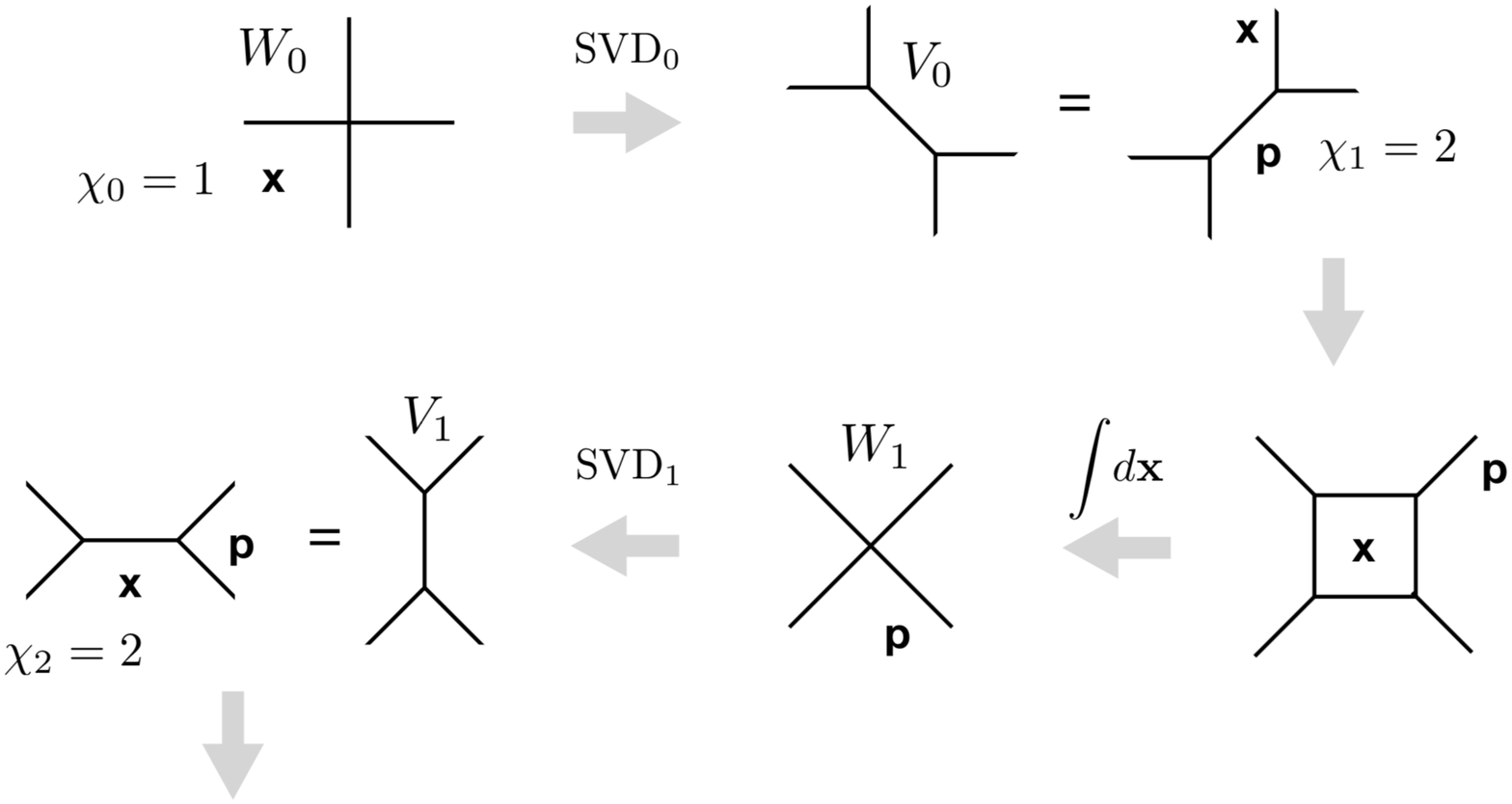}}}  \nonumber \\[-23mm]
&& \hspace{2mm}  \vcenter{\hbox{\includegraphics[width=6.5cm]{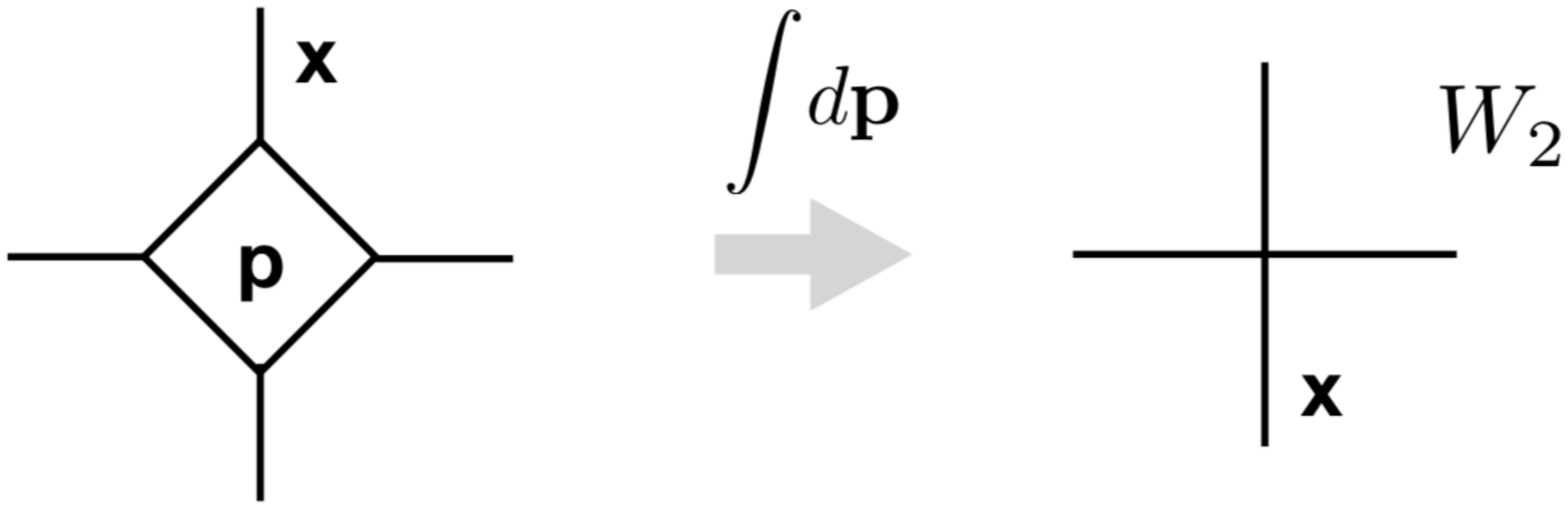}}} \hspace{-5mm} \ .
\label{TRG-steps}
\eea
\vspace*{-1.8cm}

\noindent The singular value matrices $D_{2n+1}$ turn out to have half of their diagonal entries equal to zero. The would be splitting fields associated with them can be readily discarded.
As a consequence, 
the bond dimension doubles when evolving from $x$-fields to $p$-fields but keeps constant from $p$ to $x$-fields \cite{CS19}. Therefore 
\be
\chi_{2n}=2^n \ .
\label{BD}
\ee
Notice that if we discretize each field into $\alpha$ components, the dimension of the Hilbert space per link will grow as $\alpha^{\chi_{2n}}$.

\subsection{Numerical results}

The free energy per site of the discretized free boson model
\be
f= -{1 \over N} \log Z \ ,
\ee
where $N$ is the number of lattice sites and $Z$ is the partition function \eqref{Z}, was used to test the numerical performance of this adapted TRG protocol \cite{CS19}.
The coarse graining process trades space time degrees of freedom by link variables. 
The exponential increase in the bond dimension that \eqref{BD} describes is not numerically sustainable. 
Truncation was implemented by discarding small singular values of the LR mixing matrix $B_n$, together with their associated splitting fields.
Fig.1 shows the results obtained with different maximal number of fields per link.
An average precision of $10^{-8}$ was achieved with $\chi_{\rm max}=64$ across a wide range of boson masses.
Remarkably no increase in the bond dimension was necessary to access small masses.
This allowed to evaluate the central charge of the CFT emerging in the massless limit. The value $c=1$ was reproduced to within a $10^{-5}$  error. 

\begin{figure}[h]
\begin{center}
\includegraphics[width=5.6cm]{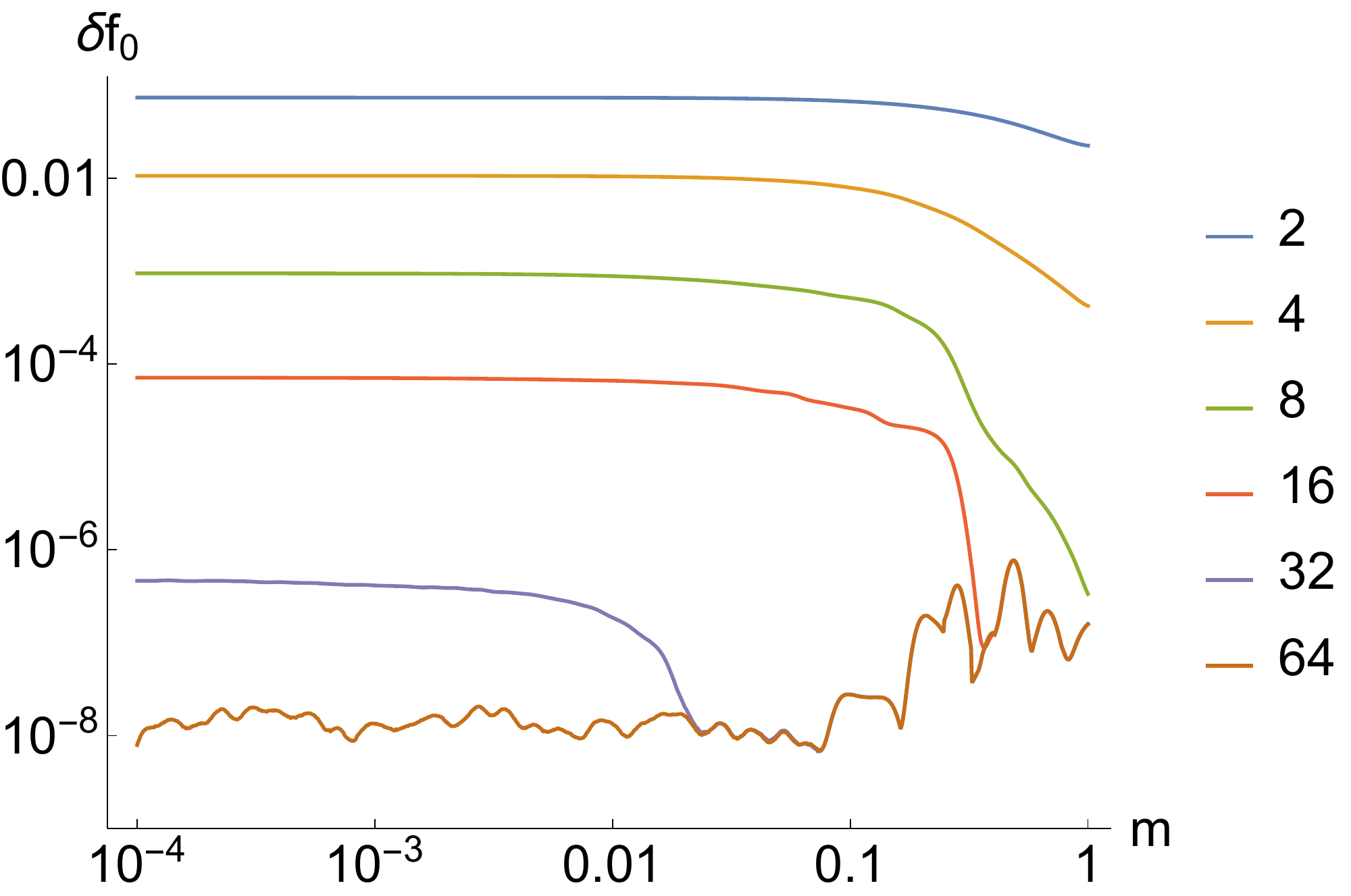}~~~
\end{center}
\vspace*{-5mm}
\caption{\label{fig:fe} Relative error in the free energy per site, $\delta f={f_{TRG}-f_{ex} \over f_{ex}}$,  as a function of the boson mass
for different maximal bond dimensions. A lattice with $N=2^{40}$ sites has been considered, such that the exact free energy $f_{ex}$ can be replaced by its $N\to \infty$ limit. 
}
\end{figure}

It was also proven that the coarse graining process attains an infrared fixed point at the right length scale set by the boson mass, even for small bond dimension.
Recall that each field encodes an infinite subset of the singular values resulting from a standard SVD, described in \eqref{39}. 
This subset contains arbitrarily small entries, which might explain why
working in terms of fields retains relevant long distance information for any bond dimension.

\section{The effective action}



The treatment of interaction will be inspired by the standard approach to quantum field theories. We will start thus reviewing some basic facts about effective field theories \cite{W75,S94}.

Let us consider a theory defined at a ultraviolet (UV) momentum scale $\Lambda$ by the Lagrangian \eqref{lagrangian}.
The partition function of the theory is given by
\be
Z=\int_{\,[0,\Lambda]} {\cal D} \phi  \; 
e^{- S[\phi] }, \quad S[\phi] =  \int d\xi {\cal L}[\phi] \ , 
\label{Z0}
\ee
where the  variables $\xi_i$ denote spacetime directions. 
Let $\Lambda' < \Lambda$ be a smaller scale. The field  $\phi$ contains {\em slow}  modes  with a momentum smaller than $\Lambda'$, 
and {\em fast}  modes with momentum in  the interval $[\Lambda', \Lambda]$. We denote them as $\phi_<$ and $\phi_>$ respectively. 
The action $S[\phi]$ can be divided into a quadratic part $S_0$, and $S_1$ which collects the interaction terms 
\be
S[\phi_<,\phi_>]= S_0[\phi_<]+ S_0[\phi_>]+\lambda S_1[\phi_<,\phi_>] \ . 
\label{Ssplit}
\ee
Quadratic cross terms between $\phi_<$ and $\phi_>$ have been disregarded because these fields are orthogonal upon integration. 
Replacing \eqref{Ssplit} into \eqref{Z0} and integrating the fast modes   yields 
\be
Z= \int {\cal D} \phi_<  \;   e^{ -S_{e\!f\!\!f}[\phi_<] }  \ .
\ee
where $S_{e\!f\!\!f}[\phi_<] = S_0[\phi_<] + S_{\rm int}[\phi_<]$ is the Wilsonian effective action at the  scale $\Lambda'$. 
The term $S_{\rm int}$ is derived from
the expectation value of $S_1$ taken with the quadratic action of the fast fields 
\barray
e^{ - S_{\rm int} [\phi_<]} & = &   \int {\cal D} \phi_>  \;  e^{-S_0[\phi_>] - \lambda   S_1[\phi_<,\phi_>] } \ .
\label{Sint}
 \earray
 
When $\lambda$ is small, the average  \eqref{Sint}  can be evaluated by means of a perturbative expansion. Its contribution to the effective action at each order in $\lambda$  is described by a sum of connected Feynman diagrams
whose internal lines are $\phi_>$ propagators. We use double lines for $\phi_>$ propagators and solid lines for $\phi_<$ external legs. Feynman diagrams will be depicted in red for avoiding confusion with the tensor network structure of the discretized model. Taking as example a $\lambda \varphi^4$ theory, some diagrams
at second order in perturbation theory are
\begin{equation}
 \vcenter{\hbox{\includegraphics[trim=4cm 14cm 4cm 14cm, clip, width=5.5cm]{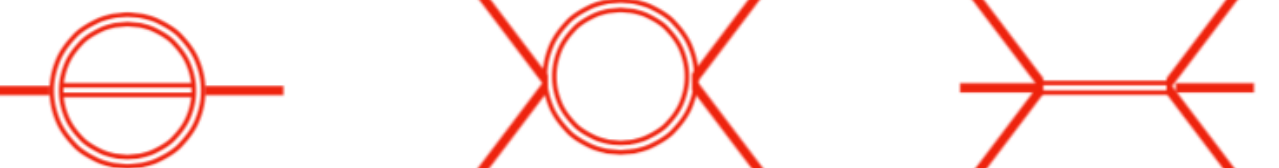}}} 
 \hspace{1mm} 
 \ .
\label{effective}
\end{equation}

\noindent The integration of high momentum modes gives rise to terms with any number of external legs.

\section{TRG for interacting quantum fields}

Guided by the previous formulation, we search for a real space RG protocol that can be implemented by local operations. Namely, we want 
the partition function to be given at each coarse graining level $n$ by
\be
Z=\int \prod_{i \in {\rm links}}  d {\bf x}_i \;\, \times \!\!\!\!\! \prod_{j \in {\rm vertices}} \!\!\! W_{n,j}({\bf x}) \ .
\ee
For concreteness we have chosen $x$-type fields to write down this general expression.
The Boltzmann weights $W_n$ should depend only on the fields living at contiguous links, as in \eqref{Wgenfree}.

The product 
\be
 \prod_{j \in {\rm vertices}} \!\!\! W_{n,j}({\bf x})  \ ,
\label{Zn}
\ee
is the real space analogue of the Wilsonian effective action, with $n$ playing the role
of the  UV cutoff $\Lambda'$. 
Preserving the structure of \eqref{Wgenfree}, we define
\be
W_n({\bf x})= e^{- {1 \over 2} {\bf x} \, M_n \, {\bf x}} \,  F_n({\bf x}) \ .
\label{WgenG}
\ee
The matrix $M_n$ encodes all $\lambda$-independent quadratic terms, being the counterpart of ${\cal L}_0[\phi_<]$.
The function $F_n$ contains the effects of interaction and therefore relates to $S_{int}[\phi_<]$.

\subsection{Real space Feynman diagrams
\label{FD}}

We analyze in this Section the new features introduced by interaction, focussing in the first TRG iteration.
The initial weights $W_0$ of the discretized interacting model  were given in \eqref{Wint}. The matrix $M_0$ is constructed from $A_0$ and $B_0$ in \eqref{A0B0}, and the function $F_0$ is 
\be
F_0({\bf x})= e^{- {\lambda \over 2} \sum_{i=1}^4 {\rm v}(x_i) } \ ,
\ee
with ${\bf x}=(x_1,..,x_4)$. The dependence of $F_0$ on $x_i$ trivially factorizes.
Anticipating the LR splitting needed to move to the next coarse graining level, we define
\be
F_0({\bf x})= 
\rho_0 \, f_0({\bf x}_L)f_0({\bf x}_R)  \ ,
\label{F0}
\ee
where 
${\bf x}_L$ labels the fields at two contiguous links entering $W_0$ and ${\bf x}_R$ those at the other two complementary links. 
A constant $\rho_0$ has been introduced for convenience as in \eqref{fr}. It is the normalization factor of the Fourier transform \eqref{fourier2}, which in this case equals $2\pi$. 

Following Section II.B for the factorization of the gaussian cross terms in $W_0$, 
the resulting cubic weights are
\vspace*{-3.6cm}
\begin{equation}
\hspace{-20mm} \vcenter{\hbox{\includegraphics[width=6cm]{cubicmain}}} \hspace{-20mm}  =\;\;\; e^{-{1\over 2} {\bf x} \, A_0 \,{\bf x} +i {\bf x} {\bf p} -{1\over 4} {\bf p}^2 } \, f_0({\bf x}) \ .
\label{cubic0int}
\end{equation}
\vspace*{-3.8cm}

\noindent The exponential prefactor reproduces \eqref{cubicfree} after taking into account that, since $B_0=\id_2$, its SVD data $U_0$ and $D_0$ are also the identity. 
Sewing together four cubic weights as shown in \eqref{TRG-steps}  and integrating out the original ${\bf x}$ fields, we obtain 
\vspace*{-3.3cm}
\begin{equation}
  \vcenter{\hbox{\includegraphics[width=5.6cm]{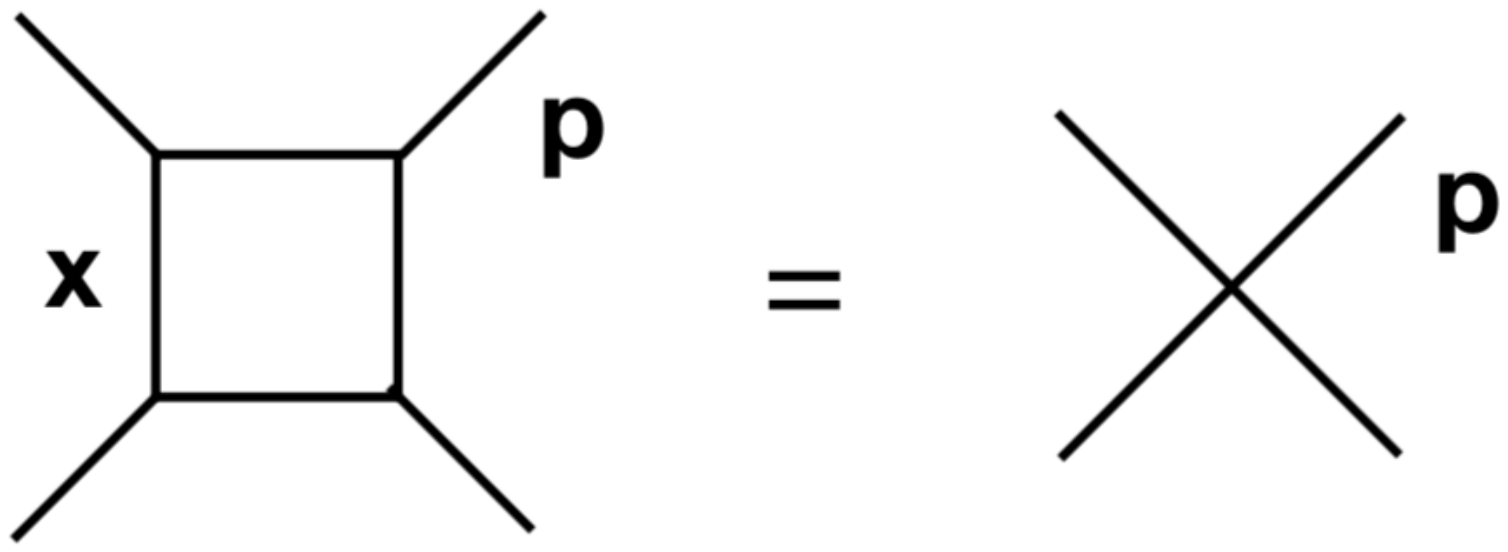}}}    \hspace{-4mm} \ , \hspace{3mm}
\label{TRGfig3}
\end{equation}
\vspace*{-3.3cm}

\noindent which stands for 
\be
W_1({\bf p})=e^{-{1 \over 4} {\bf p}^2} \! \int d {\bf x}\;  e^{-{1 \over 2} {\bf x} \, Q_0 \, {\bf x} + i {\bf x} \, C_0 \, {\bf p}} \; \prod_{i=1}^4 \, f_{0}^{(i)}({\bf x}) \ .
\label{pert}
\ee
The matrices $Q_0$ and $C_0$, of dimensions $4 \times 4$ and $4 \times 8$ respectively, are readily constructed from $A_0$ and $U_0$.
Their concrete expressions  
can be found in Appendix B. 

The gaussian kernel of \eqref{pert} can be further simplified by the change of variables
\be
{\bf x} \to {\bf x} +i Q_0^{-1} C_0 {\bf p} \ .
\label{pert2}
\ee
It leads to the structure \eqref{WgenG} for $W_1$ with
\be
M_1= {1 \over 2} \id_8 +C_0^T Q_0^{-1} C_0 \ .
\label{pert1}
\ee
The first term on the {\it rhs} reproduces the gaussian factor in front of the integral \eqref{pert}, and the second is generated by the previous shift of variables. 
The function $F_1$ encoding the effects of interaction is given by
\be
F_1({\bf p})=\int d {\bf x}\;  e^{-{1 \over 2} {\bf x} \, Q_0 \, {\bf x}} \; \prod_{i=1}^4 \, f_{0}^{(i)}({\bf x}+i Q_0^{-1} C_0 {\bf p} ) \ .
\label{F1}
\ee
This expression can be considered a local analogue of  \eqref{Sint}, with ${\bf x}$ and ${\bf p}$ identified with the fields $\phi_>$ and $\phi_<$ respectively.
We can therefore evaluate $F_1$ with the same tools used for the effective action. In particular, we might use a perturbative expansion
when the coupling constant is small.

Order by order in $\lambda$ the integral \eqref{F1} is gaussian and can be easily performed. This is done by considering all possible pairings of the $x_i$ variables and replacing 
\begin{equation}
x_i x_j \to (Q^{-1}_0)_{ij} \ ,
\label{replacement}
\end{equation} 
which is equivalent to the well known Wick's theorem. The matrix $Q_0^{-1}$ plays the role of the high momentum mode propagator.
Using this, a set of Feynman rules can be defined in terms of which to construct the real space analogue of Feynman diagrams. This is presented in Appendix C.
The logarithm of $F_1$ is given by a sum of so defined connected Feynman diagrams. 
For a $\lambda \phi^4$ theory, schematically we will have
\vspace*{-3.6cm}
\begin{equation}
\log F_1 \propto 1- \lambda  \vcenter{\hbox{\hspace{-12mm} \includegraphics[width=6.0cm]{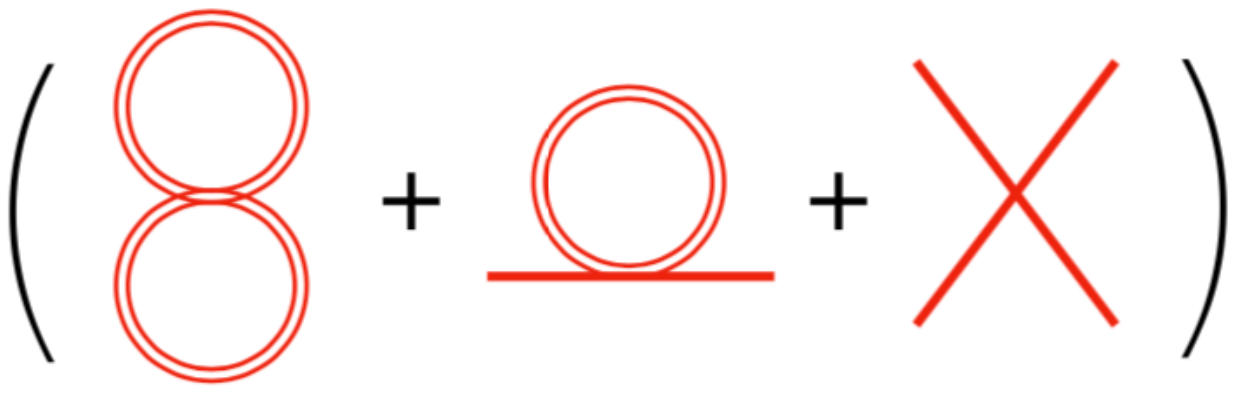}}} \hspace{-11mm} + \dots \ .
\label{Feynman1}
\end{equation}
\vspace*{-3.8cm}

\noindent  
We have followed the same graphical notation as in \eqref{effective}: a double line for the original fields ${\bf x}$, living at the four inner links on the {\it lhs} of \eqref{TRGfig3},
and a single line for the first level fields ${\bf p}$, carried by the four tilted links.

\subsection{Interacting SVD \label{gSVDS}}

The effects of interaction on the TRG protocol appear after the first coarse graining integration.
Contrary to $F_0$, the function $F_1$ does not satisfy the trivial factorization \eqref{F0}. 
This is an immediate consequence of the matrix $Q_0$ not being diagonal.

The non-factorization of $F_1$ is to be expected, since the contrary will imply that interaction has no effect on entaglement.
As a consequence of this, the LR splitting protocol has to be revisited. 
To that aim we consider
the following function of two variables, which 
serves as a simplified version of the general weights \eqref{WgenG}
\be
W(x,y) =
e^{-{b \over 2} (x-y)^2} \, F(x,y)  \ .
\label{non-gaussian}
\ee
The parameter $b$ represents a singular value of the LR mixing matrix $B$ \eqref{B}. The function $F$ is assumed to be symmetric under the exchange of $x$ and $y$, but otherwise arbitrary. This is consistent with the symmetries of the lattice model. Although $F$ does not satisfy in general  \eqref{F0}, it can be rewritten as
\be
F(x,y)= \rho \, f(x,z)\, f(y,-z) \ , \hspace{5mm} z=x-y \ ,
\label{Fff}
\ee
for some appropriate but not unique function $f$. 
A possible although not necessarily convenient choice of $f$ is
\be
f(x,z)=\sqrt{\rho^{-1} F(x,x-z)} \ .
\ee

Recall the main guidelines of the TRG protocol that we wish to design.
First, obtaining an SVD-like protocol that classifies {\it fields} 
according to an entanglement criterium. Second, keeping the coarse graining process {\it local}. This requires that no residual coupling between $x$ and $y$,
linked to the function $F$, remains after splitting. We will start by analyzing how the transformation \eqref{fourier} fails to meet the latter condition. 
Upgrading $z$ in \eqref{Fff} into an independent field, the general expression \eqref{non-gaussian} can be restated as
\be
W(x,y) =
{\rho \over 2 \pi} \int  dz  d p \, e^{\,  i p(x-y-z)-{b \over 2} z^2} \, f(x,z)f(y,-z) \ .
\label{zint}
\ee
Integrating $p$ produces a delta function that imposes $z=x-y$, and equality immediately follows. Let us invert the order of integration and
evaluate first the integral in $z$
\be
{\rho \over 2 \pi} \int  dz  \, e^{ - i pz-{b \over 2} z^2} \, f(x,z)f(y,-z) \ .
\ee
The functions $f$ are brought out of the integral by promoting them into a the differential operator
\be
{\hat f}(x,\partial_p) = f(x,z)|_{z=i \partial_p} \ .
\label{diff}
\ee
The gaussian integral can then be explicitly evaluated. We set $\rho\!=\!\sqrt{2 \pi b}$ such that numerical factors cancel, and obtain
\be
W(x,y) =
\int  dp  \, {\hat f}(x,\partial_p){\hat f}(y,-\partial_p) \, e^{-{1 \over 2b} p^2} \ .
\ee
When $f$ is independent of $z$, as in \eqref{F0}, 
the standard Fourier transform is recovered. In the general case, the fact that both differential operators act on the same gaussian factor hinders factorization.

This problem has a simple solution. We may double the splitting fields such that each differential operator acts on a different gaussian factor.
This is achieved by considering
\be
F(x,y)={\tilde \rho}
\,f(x,z^1)\, f(y,-z^2) \ , \hspace{5mm} z^{1,2}=x-y \ ,
\label{Fff12}
\ee
with ${\tilde \rho}=\rho^2/2$.
We upgrade again $z^{1,2}$ into independent variables with the help of two Fourier fields $p^{1,2}$.
Using this trick \eqref{non-gaussian} can be factorized as follows
\be
W(x,y) =
\int \!\! d\!p^1\! d\! p^2 \; V(x;p^1\!\!,p^2) \;V^\ast(y;p^2\!\!,p^1)  \ ,
\ee
\vspace*{-4mm}
\noindent with
\be
V(x;p^1\!\!,p^2) = e^{\, ix(\!p^1\!+p^2) } \,  \Big( {\hat f}(x,{\partial}_{p^1})  \, e^{-{(p^1)^2 \over b}} \Big) \ .
\label{Vsplit}
\ee
Notice that $V$ depends on both splitting fields, albeit in an asymmetrical manner. 

\subsection{Feynman diagrams and SVD}

We have seen that the function $F_1$ encoding the interaction effects of the level one Boltzmann weights, admits a diagrammatic expansion. We analyze here how this feature behaves under the factorization protocol.

Recall that the first TRG iteration exchanges the roles of the $x$ and $p$ variables. Following \eqref{Fff12}, we define
\be  
F_1({\bf p}_L, {\bf p}_R)= {\tilde \rho}_1 \, f_1({\bf p}_L,{\bf z}^1 ) \, f_1({\bf p}_R,-{\bf z}^2 ) \ , 
\label{Fnff}
\ee
with ${\tilde \rho}_1$ an appropriate constant and
\be
{\bf z}^1={\bf z}^2={\bf p}_L-{\bf p}_R \ .
\label{z12}
\ee
In order to understand the structure of $f_1$, we focus on a simple example.
Let us consider a contribution to $F_1$ of the form
\be
e^{- \lambda  ({\bf p}_L \cdot {\bf p}_R )^2} \ . 
\ee
The argument of the exponential represents a connected Feynman diagram with four legs contributing at first order in perturbation theory. It has the same structure as
the last graph in \eqref{Feynman1}.
The splitting of the previous term according to \eqref{Fnff} is
\be
e^{- {\lambda \over 2}   \, [{\bf p}_L \cdot ({\bf p}_L- {\bf z}^1)]^2} \; e^{- {\lambda \over 2}  \, [{\bf p}_R\cdot ({\bf p}_R+{\bf z}^2 ) ]^2} \ .
\label{factorex}
\ee
We observe that the four leg diagram is inherited by both factors. The coefficient weighting the diagram has been modified such that the original term is recovered when substituting \eqref{z12}. 
The undesired dependence on the fields ${\bf p}_{L,R}$ has been replaced with the help of the auxiliary variables ${\bf z}^{1,2}$.

The previous example illustrates a general conclusion. Connected Feynman diagrams in the expansion of $F_1$ reappear in $f_1$ with their topological structure unchanged.
Therefore, since the former has a diagrammatic expansion so it does the latter. 
As shown in \eqref{Feynman1} the logarithm of $F_1$ only contains connected Feynman diagrams. Expression \eqref{factorex} makes clear that this property is also inherited by $f_1$. 
By induction, these statements hold at all coarse graining levels.

\section{Interaction and entanglement}

The semi-analytical character of the coarse graining protocol that we have designed allows to obtain interesting results on the interplay between interaction and entanglement already before addressing its numerical implementation.
That will be the subject of this Section.

\subsection{A tale of two fields}

We are ready to factorize the level one Boltzmann weights.
As explained above, two sets of splitting fields ${\bf x}^{1,2}$ are necessary. 
Upper indices will always make reference to the splitting process.
The resulting cubic weight $V_1({\bf p};{\bf x}^1\!\!,{\bf x}^2)$ is
\vspace*{-3.7cm}
\begin{eqnarray}
& \hspace{-8mm} \vcenter{\hbox{\hspace{-20mm} \includegraphics[width=6.3cm]{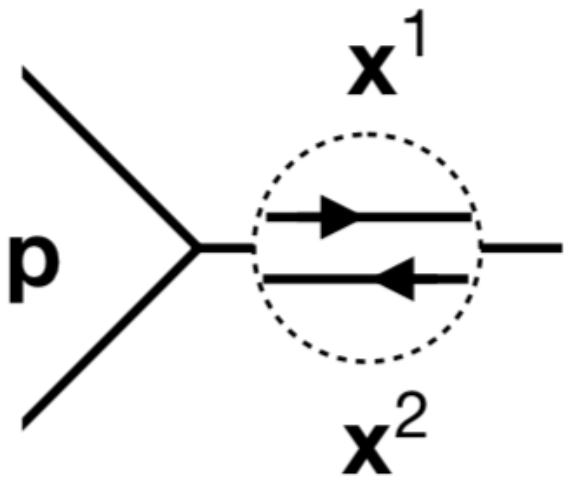}}} \hspace{-19mm} \label{V11bis} &
= \;\; e^{-{1 \over 2} {\bf p} \,A_1 {\bf p}+ i \,{\bf p} \, U_1 ({\bf x}^1+{\bf x}^2)}\,  
\\[-39mm]
&&  \;\;\;\;\;\; \times \; \left(  {\hat f}_1 \! \big({\bf p},{\partial}_{{\bf x}^1}\big) \, e^{-{\bf x}^1\!  D_1^{-1}  {\bf x}^1} \right) \ . \nonumber
\end{eqnarray}
\vspace*{-.1cm}

\noindent Generalizing \eqref{diff}, the differential operator ${\hat f}_1$ is given by 
\be
{\hat f}_1 \! \big({\bf p},{\partial}_{{\bf x}^1}\big) = f_1 ({\bf p},{\bf z})|_{{\bf z} =  i \, U_1 \partial_{{\bf x}^1}} \ .
\label{f1}
\ee
The matrices $A_1$, $U_1$ and $D_1$
governing the gaussian terms are those already present in the free case \eqref{cubicfree}.

The arrows in the magnified link stress the different roles that the fields ${\bf x}^{1,2}$ play in the cubic weight.  
This is better understood by considering the linear combinations
\be
{\bar{\bf x}}= {\bf x}^1\!+{\bf x}^2 \ , \hspace{5mm} {\hat {\bf x}}= {\bf x}^1\!-{\bf x}^2 \ .
\label{pm}
\ee
We will refer to them as even and odd variables respectively.
The $\lambda$-independent coupling of ${\bf p}$ with the splitting fields only involves the even variables $\bar {\bf x}$. In the absence of interaction the odd variables 
$\hat {\bf x}$ completely decouple, since multiplying contributions from the L and R cubic weights we have
\be
e^{-{\bf x}^1\!  D_1^{-1}  {\bf x}^1-{\bf x}^2  D_1^{-1}  {\bf x}^2}=e^{-{1 \over 2}  {\bar {\bf x}} \, D_1^{-1}  {\bar {\bf x}}-{1 \over 2} {\hat {\bf x}} \, D_1^{-1}  {\hat {\bf x}}} \ .
\label{nocoupling}
\ee
Therefore ${\bar {\bf x}}$ is the counterpart of the splitting fields already present in the gaussian SVD, while $\hat {\bf x}$ 
allows to treat locally the new entanglement ties
created by interaction. Without interaction the latter become trivial.

Sewing four cubic weights \eqref{V11bis} and following the steps presented in Section IV.A for the elimination of ${\bf p}$, we obtain the second level Boltzmann weights 
\be
W_2({\bf x})= e^{-{1 \over 2} {\bf x} \, M_2 {\bf x} } \, F_2 ({\bf x}) \ .
\ee
Here ${\bf x}$ stands for both splitting fields  $({\bar {\bf x}}, {\hat {\bf x}} )$. Recall that the gaussian prefactor only contains leading terms in the pertubative expansion, namely ${\cal O}(\lambda^0)$.
Its exponent is given by
\be
{\bf x} \; M_2 \,{\bf x} =
{\bar {\bf x}}\Big( {1 \over 2} S_1 + C_1^T Q_1^{-1} C_1 \Big){\bar {\bf x}} +{1 \over 2} {\hat {\bf x}} \,S_1{\hat {\bf x}} \ ,
\label{M2}
\ee
with $S_1 =\id_4 \otimes D_1^{-1}$ diagonal, and $Q_1$ and $C_1$ constructed out of $A_1$ and $U_1$ respectively as explained in Appendix B. 
The expression in parenthesis 
is the same combination governing the level one Boltzmann weights in the absence of interaction. This shows the direct 
connection between ${\bar {\bf x}}$ and the fields already present in the free network.
We have disregarded quadratic terms between even and odd variables. 
Indeed, relation \eqref{nocoupling} implies that they trivially cancel in the product of Boltzmann weights \eqref{Zn} building up the partition function.

The effect of interaction is encoded in the function
\be
F_2({\bf x})= 
 e^{\,{\bf u} \, S_1 {\bf u}}\, {\hat {\cal F}}_2({\bar {\bf x}},\partial_{{\bf u}} ) \; e^{-{\bf u} \, S_1 {\bf u}} \, \big|_{{\bf u}={\bf x}^1} \ .
\label{hermite}
\ee
A variable ${\bf u}$ has been introduced in order to unambiguously define the action of the derivatives inherited from ${\hat f}_1$.
The differential operator ${\hat {\cal F}}_2({\bar {\bf x}},\partial_{{\bf u}} )$ is given by
\be
 \int d{\bf p} \; e^{-{1 \over 2} {\bf p} \, Q_1 {\bf p}}  \prod_{i=1}^4 \, {\hat f}_1^{(i)}\! \big({\bf p} + i Q_1^{-1} C_1 {\bar {\bf x}},\partial_{{\bf u}} \big) \ .
\label{F2} 
\ee
This integral is the level one equivalent of \eqref{F1}. 
The expression \eqref{hermite} may seem unconventional. It is however reminiscent of the Hermite polynomials
\be
H_n(x)= (-1)^n \,e^{\,x^2} {d^n \over dx^n} e^{-x^2} \ .
\label{her}
\ee
Each monomial $x^k$ in $H_n$ is to be related with a Feynman diagram with $k$ external legs of type ${\bf u}$. 
This introduces an additional structure in the set of Feynman diagrams, whose relevance we discuss below.

\subsection{General structure}

The main feature of our protocol is the introduction of new fields representing the effect of interaction on entanglement. 
They first appear at the second coarse graining level, corresponding to the odd combination of the splitting fields.

We recall schematically the leading order structure of the even and odd fields, ${\bar {\bf x}}$ and ${\hat {\bf x}}$ respectively, using the decomposition \eqref{Mfree} of $M_2$
\be
A_2   = \begin{pmatrix} A_{2{\bar {\bf x}}} & 0 \\ 0 & A_{2{\hat {\bf x}}} \end{pmatrix} \ , \;\;\;\;\;\;\;\;\; 
B_2   = \begin{pmatrix} B_{2{\bar {\bf x}}} & 0 \\ 0 & 0 \end{pmatrix} \ .
\label{M22}
\ee
The decoupling between even and odd variables, together with the vanishing of the odd mixing matrix $B_{2{\hat {\bf x}}}$, shows the trivial nature of ${\hat {\bf x}}$ at leading order ${\cal O}(\lambda^0)$.
Namely, these fields just represent decoupled degrees of freedom living at each link. However they contribute in general to LR mixing through the function $F_2$, and require an own set of splitting fields to continue the coarse graining process. 

In the simplified splitting problem \eqref{non-gaussian}, the vanishing of $B_{2{\hat {\bf x}}}$ is equivalent to setting $b=0$.
It is thus a question whether the factorization technique of Section III.B applies in this limiting situation, which moreover will be recurrent in subsequent iterations.
In order to clarify this point we send $b$ to zero in \eqref{Vsplit}, after renaming the splitting fields as $q^{1,2}$. The product of the gaussian 
exponential and its associated  normalization factor $\rho$ results then in a delta function
\be
\lim_{b \to 0}  {1 \over \sqrt{\pi b}} e^{-{1 \over b} (q^1)^2} = \delta(q^1) \ .
\label{delta}
\ee
Namely, the consequence of a vanishing $b$ is the collapse into a delta function of the probability distribution of $q^{1}$.
In the free case, a delta function implies that the associated field can be trivially removed from the coarse graining process. In the presence of interaction, this does not hold
anymore. 

We come back to the factorization of $W_2({\bf x})$. 
The splitting fields for the even and odd variables
${\bar {\bf x}}$ and ${\hat {\bf x}}$ will be denoted by ${\bf p}$ and ${\bf q}$ respectively. 
As just explained, the probability distribution of fields ${\bf q}$ collapses into a delta function.
The level two cubic weights 
are then
\vspace*{-3.5cm}
\begin{eqnarray}
&& \vcenter{\hbox{\hspace{-28mm} \includegraphics[width=6.0cm]{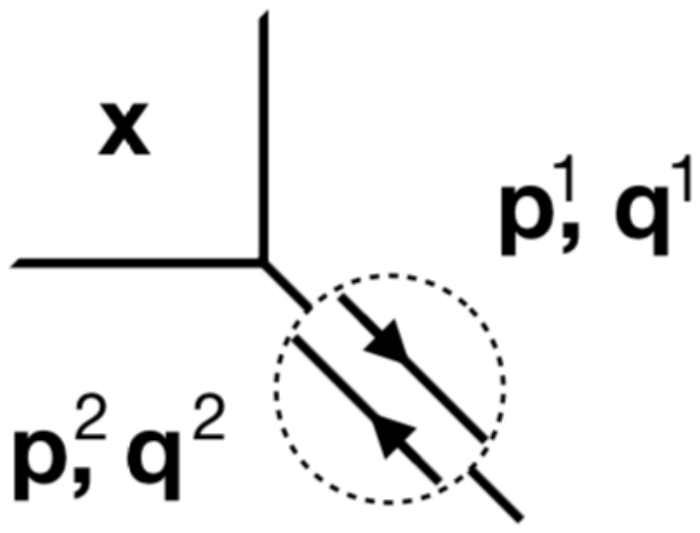}}} \hspace{-17mm}
 = \;\;\;e^{-{1 \over 2} {\bf x} \,A_2 {\bf x} \,+\, i \,{\bar {\bf x}}\, U_2 \, {\bar {\bf p}}\,  + \, i \,{\hat {\bf x}}\, {\bar {\bf q}}}  \, \times \label{cubicgen} \\[-38mm]
&& \hspace{17mm}\; {\hat f}_2 \big( {\bf x},{\partial}_{{\bf p}^1}\!, {\partial}_{{\bf q}^1} \!\big) \!\! \left(\! e^{-{\bf p}^1\!  D_2^{-1}  \, {\bf p}^1} \delta ({\bf q}^1 \!) \! \right) \ , \nonumber
\label{qq}
\end{eqnarray}

\noindent with $U_2$ and $D_2$ the SVD data of $B_{2{\bar {\bf x}}}$.
The differential operator ${\hat f}_2$ induces derivatives of 
$ \delta ({\bf q}^1 \!)$. The delta derivatives
can not be evaluated right away. This has
to be postponed to the coarse graining integral
that eliminates ${\bf q}$. It will be done using integration by parts, which is equivalent to work at the level of the Hermite polynomials
\eqref{her} instead of dealing with individual Feynman diagrams.
In this way the fields ${\bf q}$, in spite of their singular distribution function, provide a regular and well defined contribution to the propagation of entanglement.

Due to the need of doubling the splitting fields, the bond dimension of the TRG network with interaction increases twice as fast an in the free case.
However the different roles of the even and odd combinations in perturbation theory induce a clear distinction between fields.
A reduced set of them will be in direct correspondence with the fields already present in the free network, inheriting from them their leading order structure.
This set is given by
\vspace*{-2.2cm}
\begin{equation}
 \vcenter{\hbox{\hspace{-20mm} \includegraphics[width=7.3cm]{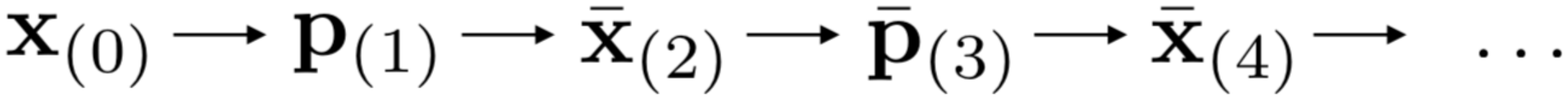}}} \hspace{-10mm}
\label{chain} 
\end{equation}
\vspace*{-2.4cm}

\noindent  where the index in parenthesis labels the coarse graining level. 
Hence ${\bar {\bf x}}_{(2)}$ refers to the corresponding fields in \eqref{M2} and ${\bar {\bf p}}_{(3)}$ to those in \eqref{cubicgen}.
The dots stand for the even combination of splitting fields associated with the previous term in the chain.

The largest set of fields represents entanglement ties created by interaction. At each coarse graining level a new family of them is seeded by the odd counterpart
of the variables \eqref{chain}
\vspace*{-2.4cm}
\begin{equation}
 \vcenter{\hbox{\hspace{-17mm} \includegraphics[width=5.3cm]{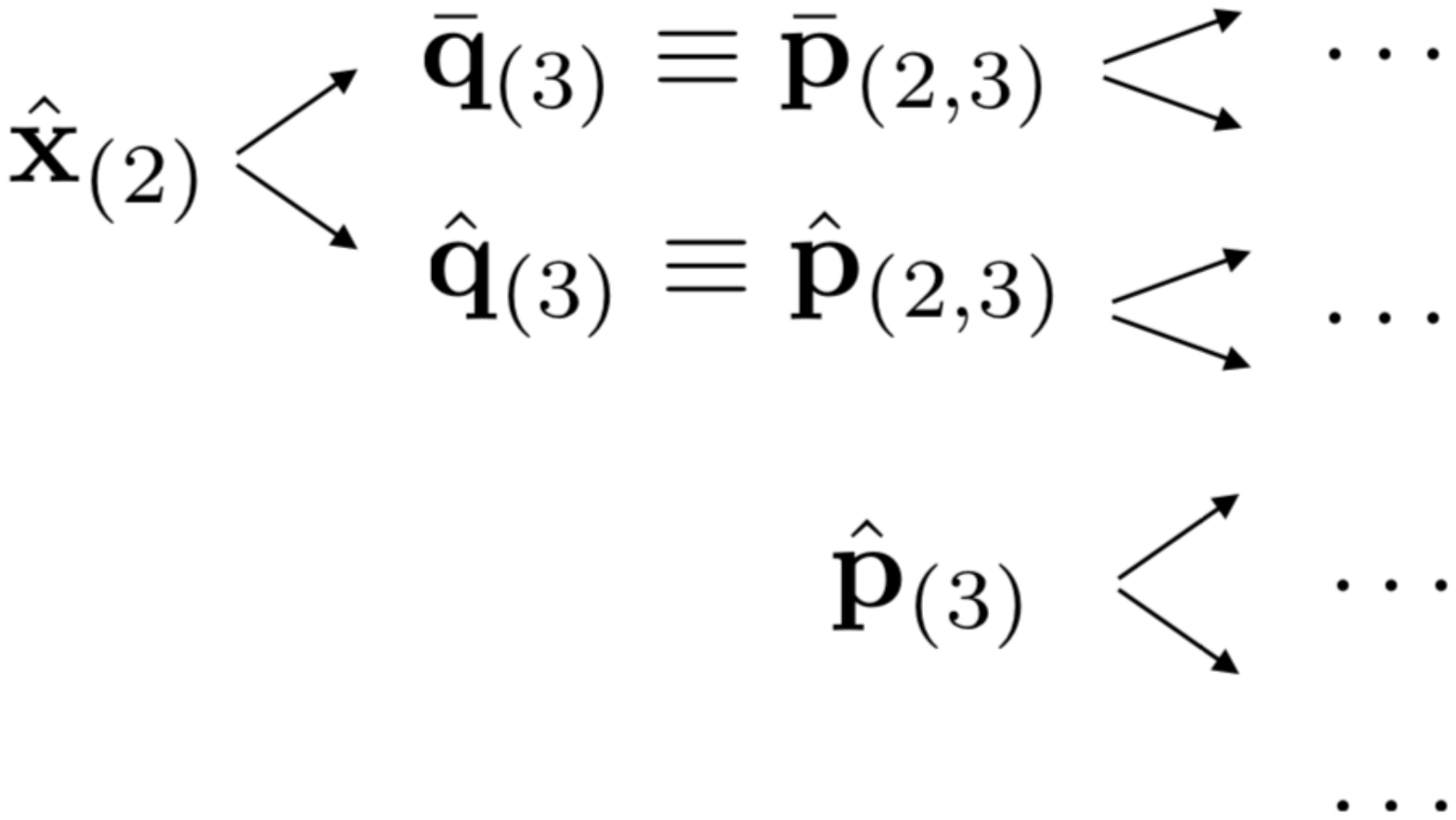}}} \hspace{-10mm}
\label{newfields} 
\end{equation}
\vspace*{-2.6cm}

\noindent In order to stress the structure organizing this set of fields,
we have renamed ${\bf q}$ in \eqref{cubicgen} as ${\bf p}_{(2,3)}$. 
All these fields have a trivial leading order structure, as ${\hat {\bf x}}_{(2)}$ in \eqref{M2}. Many of them will have a singular 
distribution function, as it is the case of ${\bf p}_{(2,3)}$, but this does not represent a harm for the TRG protocol.

\subsection{A perturbative bound on $\chi$}

In the previous section we have not made reference to the order at which we want the perturbative expansion to stop.
This is however an important information for the TRG protocol, which sets a limit to the propagation of entanglement mediated by odd variables and
results in a reduction of the large set of extra fields \eqref{newfields}.

For $n \geq 2$ it is best to analize the effects of the interaction in terms of  
a differential operator ${\hat {\cal F}}_n({\bar {\bf x}},\partial_{\bf u} )$, generalizing ${\hat {\cal F}}_2$ in \eqref{hermite} and defined by an integral as in \eqref{F2}
\vspace*{-3.1cm}
\begin{equation}
\hspace{-5mm} {\hat {\cal F}}_n({\bar {\bf x}},\partial_{\bf u} )\;\; = 
\vcenter{ \hbox{\hspace{-16mm} \includegraphics[width=6.1cm]{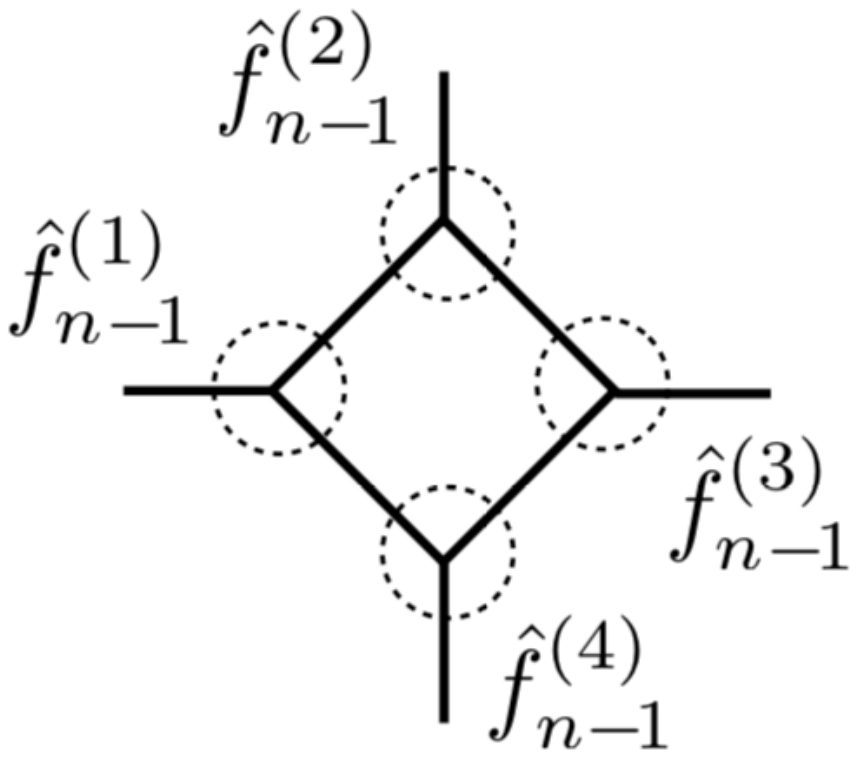}}} \hspace{-15mm}  \ . \label{Fgraph1}
\end{equation}
\vspace*{-3.4cm}

\noindent 
Recall that ${\bf u}$ is an auxiliary variable which, after the evaluation of the derivatives, should be replaced by ${\bf x}^1$ introducing therefore a dependence on the odd variables.
The operator ${\hat {\cal F}}_n$ admits an expansion in Feynman graphs, where the graphs's external legs can carry components of the vector ${\bar {\bf x}}$ or the derivative $\partial_{\bf u}$.
Let us assume that in order to move to the next coarse graining level we factorize \eqref{Fgraph1} along the axis $i: (1,2)_L$-$(3,4)_R$.
A generic connected diagram in the expansion of ${\hat {\cal F}}_n$ has the structure
\vspace*{-4.0cm}
\begin{equation}
\vcenter{\hbox{\hspace{-14mm} \includegraphics[width=6.6cm]{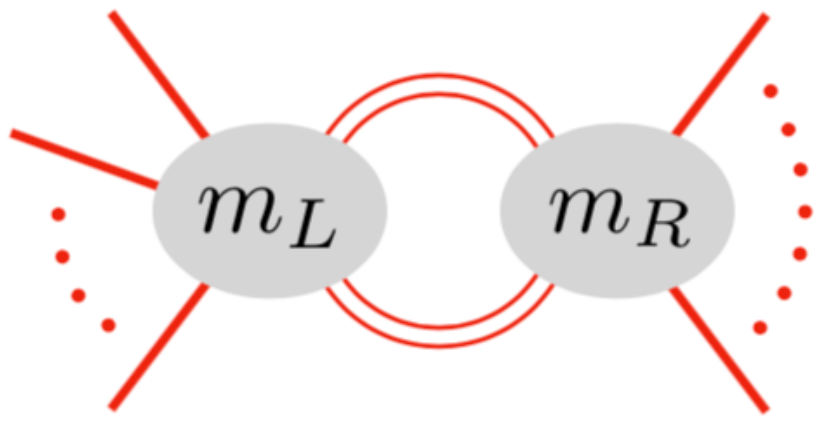}}} \hspace{-10mm} 
\label{LRgraph1}
\end{equation}
\vspace*{-4.0cm}

\noindent 
The subgraph on the left originates from ${\hat f}_{n-1}^{(1,2)}$ and contributes $m_L$ powers of the coupling constant.
A corresponding statement holds for the subgraph on the right.
After introducing the necessary splitting fields, the previous graph will be inherited by the cubic weights resulting from the factorization of \eqref{Fgraph1} as described in Section IV.C.
Appendix D proves that
splitting fields associated with the odd variables ${\hat {\bf x}}$, such as ${\bf q}^{1,2}$ in \eqref{qq}, are only required in diagrams satisfying
\be
m_L, m_R \geq 1 \ , \;\;\;\;\;\; m_L+m_R \leq N \ ,
\label{oddfact1}
\ee
with $N$ the order at which to cut the perturbative expansion.

Clearly these conditions cannot be met at first order in perturbation theory.
Although doubling the splitting fields is necessary, at first order in $\lambda$ the odd combinations represent short range entanglement and their effect extinguishes within one iteration
\vspace*{-2.1cm}
\begin{equation}
 \vcenter{\hbox{\hspace{-15mm} \includegraphics[width=7.3cm]{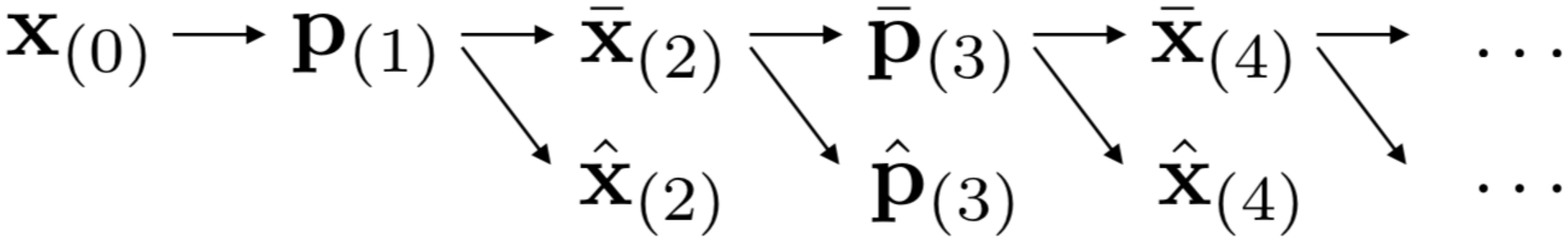}}} \hspace{-10mm} \label{onechain}  \hspace{3mm} \ .
\end{equation}
\vspace*{-2.3cm}

\noindent  The upper row contains the fields \eqref{chain} with a counterpart in the free network. 
From the set \eqref{newfields} only the first item in each tree is relevant. This drastic simplification is consistent with the rather trivial nature of quantum corrections at order $\lambda$.

At second order in perturbation theory \eqref{oddfact1} implies
\be
m_L=m_R=1 \ .
\label{secondO}
\ee
Thus if a Feynman diagram contributes to odd variable mixing in the $n$-{\it th} iteration, it will not do it at any subsequent one.
Indeed, as an individual diagram it fails to satisfy the first condition \eqref{oddfact1}, while as part of a larger diagram it will be the second which fails.
The fields involved at second order include \eqref{onechain}, plus secondary chains with that same structure stemming from each element of the second row.
Namely the second level odd fields, such as ${\hat {\bf p}}_{(2,3)}$ in \eqref{newfields}, get extinguished within one iteration at second order in $\lambda$.
It is straightforward to generalize this reasoning to order $N$, with the result that at larger $N$ more ramifications in \eqref{newfields} become relevant.
Equivalently, high orders in perturbation theory require a growing number of extra fields to represent locally the effects of interaction. 
We consider this property a consistency check of our proposal.

We shall end with a comment on the fields \eqref{chain}. It was mentioned in Section II.A that before truncation kicks in, the singular values matrices $D_{2n+1}$ have half of their diagonal entries equal to zero.
The probability distribution of the associated fields is thus a delta function. Although in the free case these fields are trivially discarded,
in the presence of interaction some of them become relevant and need to be kept. We show in Appendix E that this only happens when conditions  \eqref{oddfact1} are met.
Since these extra fields 
behave with respect to the perturbative order as the odd variables do, 
we will not discuss them further.

\vspace*{-1.mm}

\section{Numerical results}

\vspace*{-1.mm}

In the previous sections we have extended the TRG protocol based on local operations formulated in terms of fields to include interactions.
We will start here the analysis of its numerical performance using as benchmark a $\lambda \phi^4$ theory. This theory has been studied with
tensor networks techniques which imply the discretization of the field variables in \cite{Sh12,K19,V19,D20,V21} 
and with continuous tensor networks techniques in \cite{T21a,T21b}.

\vspace*{-1.mm}

\subsection{Truncation}

The last question that needs to be addressed before implementing the TRG is truncation.
In the free case truncation consists in discarding small singular values of the matrix $B_n$. 
As \eqref{delta} shows, this is equivalent to approximate the probability distribution of the splitting fields related with small singular values by a delta function.

Perturbation theory treats in an asymmetrical way quadratic terms and interaction, giving rise to a potential problem. Truncation is only justified 
for fields with small LR cross terms from {\it both} the leading order gaussian prefactor and the Feynman diagrams. 
This only represents a problem for fields contributing non trivially at leading and higher order in $\lambda$, namely, the set \eqref{chain}.
These two conditions are a priori not related, since fields with a delta function distribution can 
yet have a relevant contribution to Feynman diagrams via delta derivatives.
Checking their compatibility is the first issue in the numerical study of the TRG.
We will work below at first order in perturbation theory to simplify the analysis by focussing as much as possible on the set \eqref{chain}.

When the TRG protocol is applied at first order in $\lambda$, there is one-to-one correspondence between the leading order gaussian structure of the even fields, and their counterpart in the
free network. See Appendix E for details.
Since the odd fields do not contribute to the propagation of entanglement, they play a secondary role. 
Motivated by this we will apply the term bond dimension to the number of fields in \eqref{chain}, which without truncation is given by \eqref{BD}. As for the criterium of truncation, 
we will adopt the simplest choice and base it again on the singular values of $B_n$.

\subsection{Partition function}

We present the results on the numerical evaluation of the free energy per site of a $\lambda \phi^4$ theory 
\be 
- {1 \over N} \log Z = f_0\ + \lambda f_1 + {\cal O}(\lambda^2) \ .
\ee
Fig.2 shows the relative error in the first order correction $f_1$
for several maximal bond dimensions and masses. The relative error of the leading term $f_0$ is also included for comparison.
A remarkable feature of the TRG protocol is the capacity to correctly estimate $f_0$ in the massless limit without varying the bond dimension.
This property however does not extend to $f_1$, which requires a sustained increase in the bond dimension to keep the precision as
the mass lowers. On the other hand, the convergence with $\chi_{\rm max}$ appears to be faster for $f_1$ than for $f_0$.
Notice that the precision in $f_1$  improves on that of $f_0$ down to  $m\approx0.06$ with just $\chi_{\rm max}=20$.

\begin{figure}[h]
\begin{center}
\includegraphics[width=6.2cm]{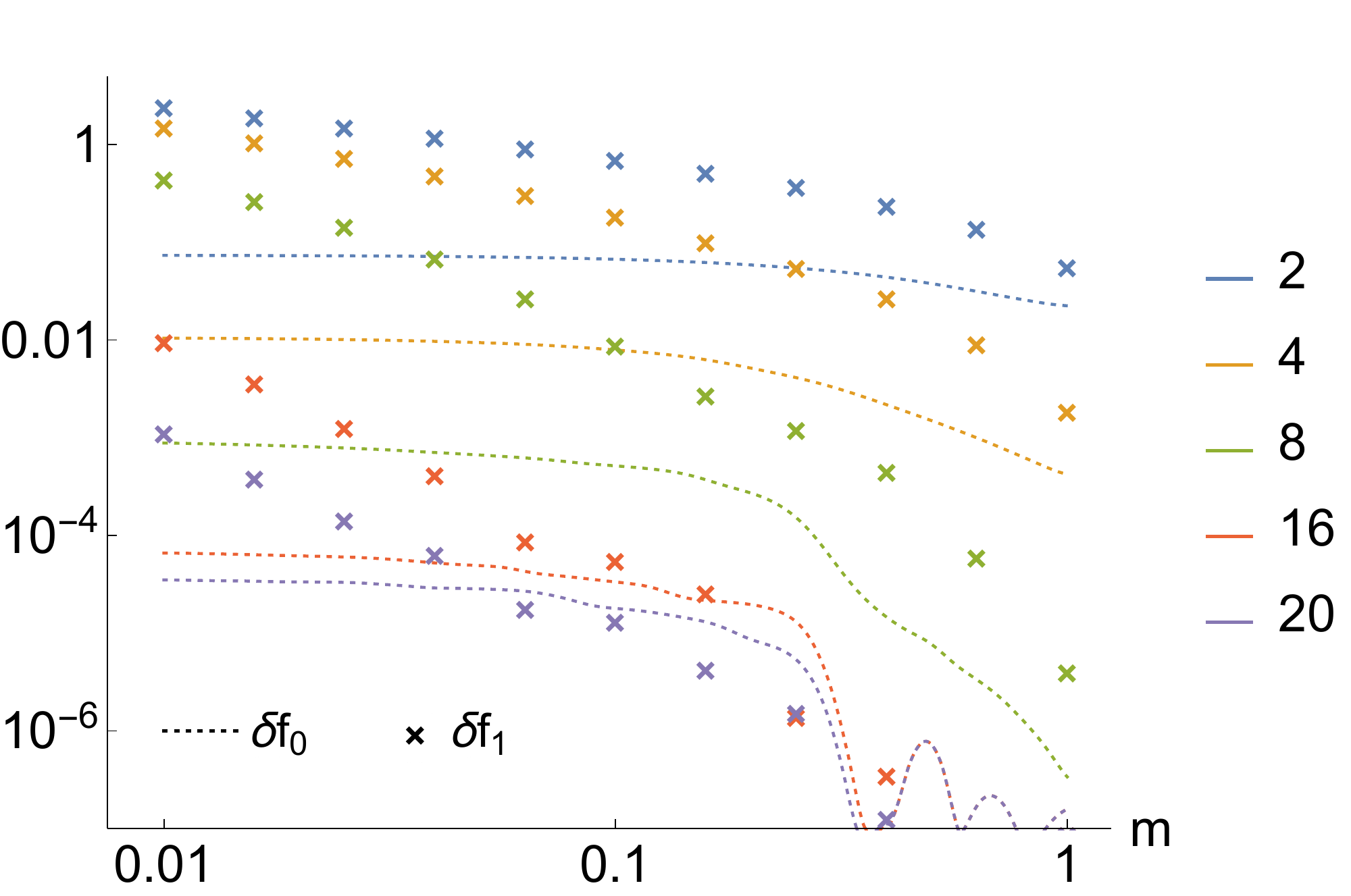}~~~
\end{center}
\vspace*{-5mm}
\caption{\label{fig:fe0} Relative errors $\delta f_0$ (dotted lines) and $\delta f_1$ (crosses) in the free energy per site as a function of the mass and the bond dimension, in a
 lattice of size $N=2^{40}$.} 
\end{figure}

It is important to quantify the convergence of the TRG with the bond dimension. The error in the leading term of the free energy at fixed mass follows a 
power law, $( \chi_{\rm max})^{-\alpha}$. This is observed in Fig.3a, where
choosing as example $m=0.1$ and setting $\alpha=3.7$ we have obtained a very good fit of $\delta f_0$. 
The behaviour of $\delta f_1$ 
is plotted in Fig.3b at the same mass. 
The error for small bond dimension, although larger than that of the leading term, decreases exponentially fast.
As the precision improves, the dependence on $\chi_{\rm max}$ turns into a power law with exponent $\alpha=7.3$.
This confirms the faster
convergence of $\delta f_1$ across all bond dimensions. 
\begin{figure}[h]
\begin{center}
\includegraphics[width=4cm]{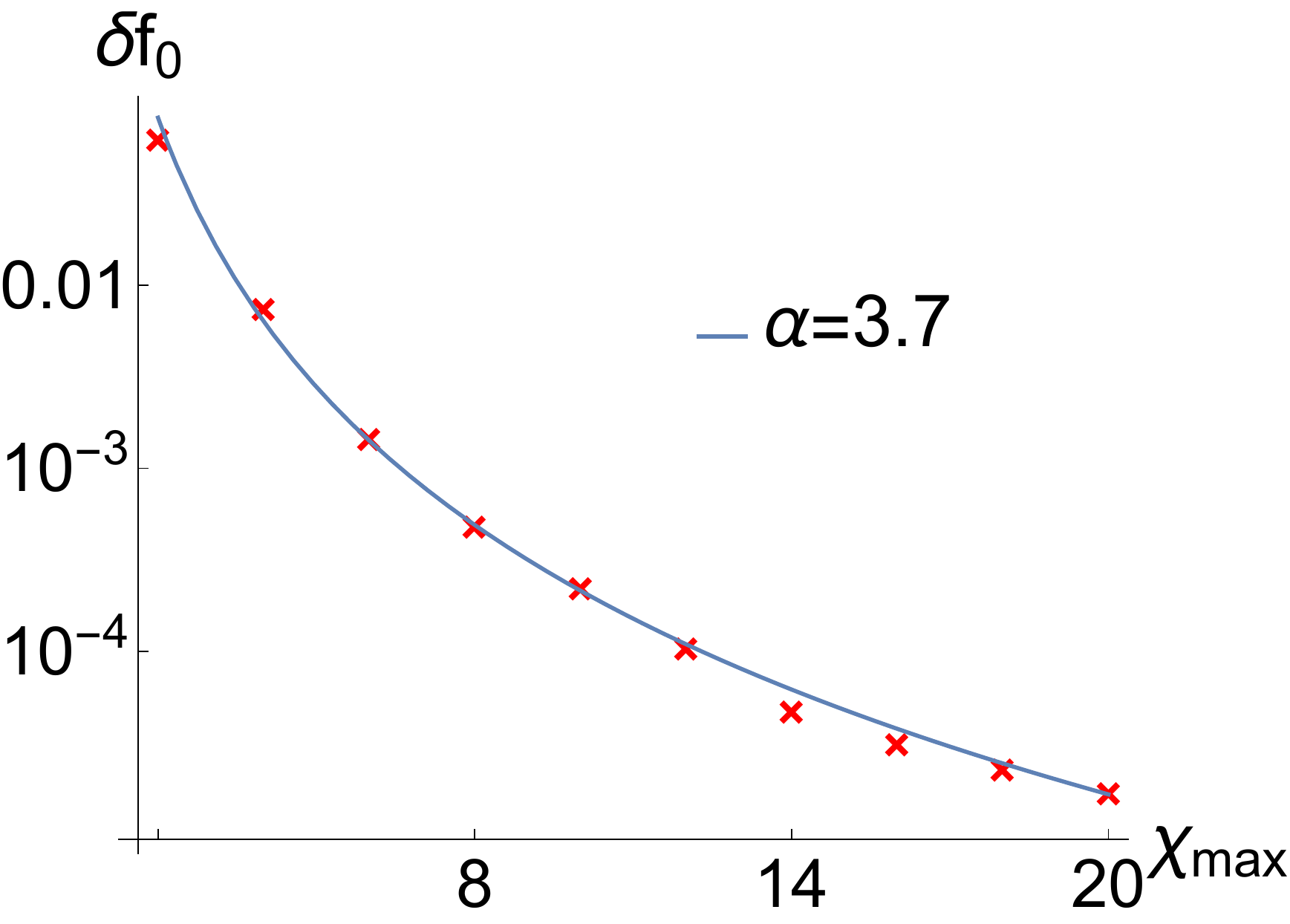}\includegraphics[width=4cm]{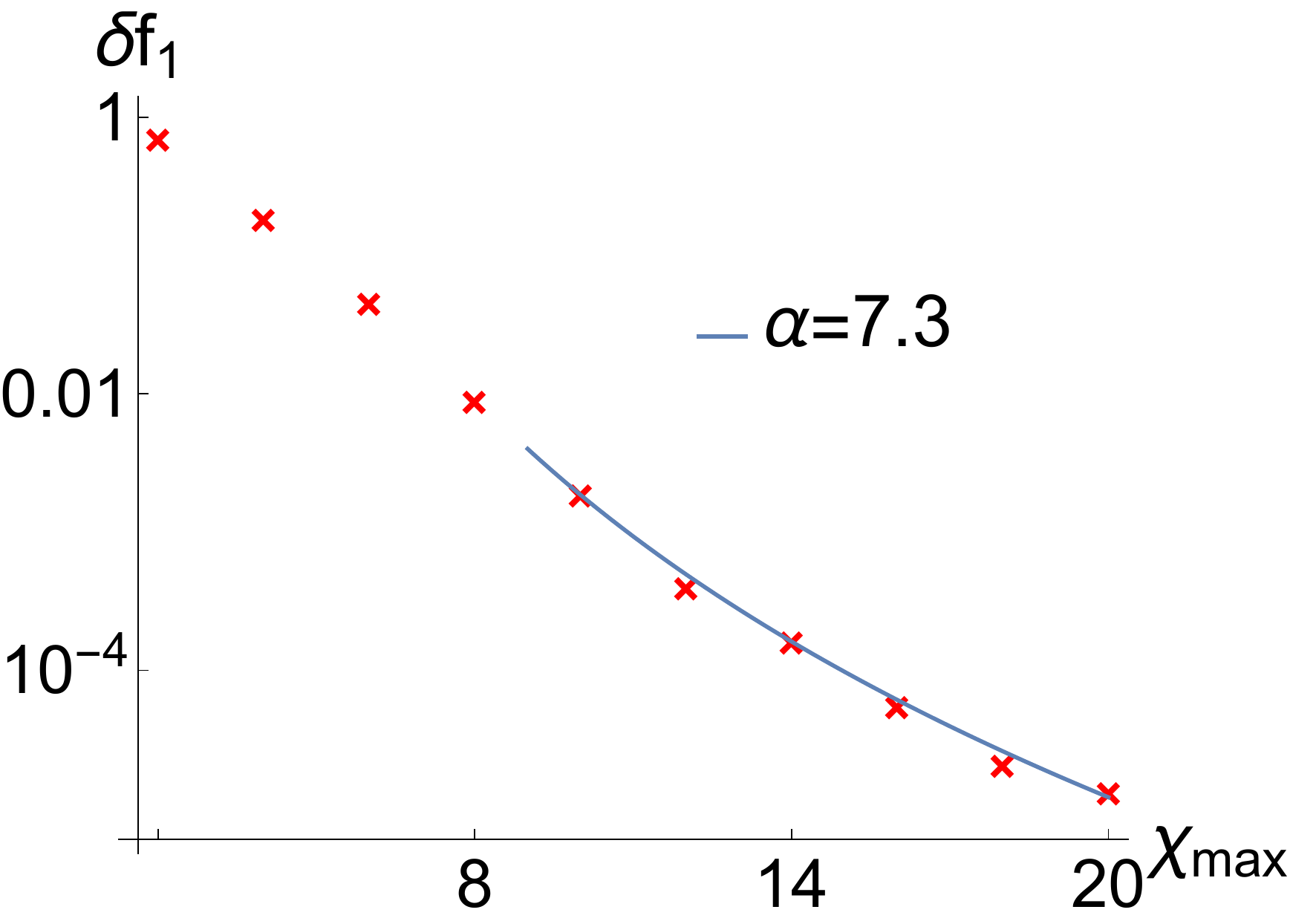}
\end{center}
\vspace*{-5mm}
\caption{\label{fig:fe1} Relative errors a) $\delta f_0$ and b) $\delta f_1$ as a function of the bond dimension at $m=0.1$. The blue lines represent a power law fit $(\chi_{\rm max})^{-\alpha}$. }
\end{figure}
A similar pattern holds at other values of the mass. The smaller the mass, the bigger the error at fixed bond dimension and the larger the range of $\chi_{\rm max}$ were
convergence is exponential. 

In the same way that the singular values of $B_n$ describe the entanglement hierarchy at leading order, new quantities are needed to characterize how interaction modifies it.
To this aim we will use the cubic weights, $V_n$. After the maximal bond dimension is reached, they result from the factorization of the Boltzmann weights plus a truncation guided by the matrix $B_n$
\vspace*{-2.2cm}
\begin{equation}
\hspace{-7mm}\vcenter{\hbox{\includegraphics[width=8.5cm]{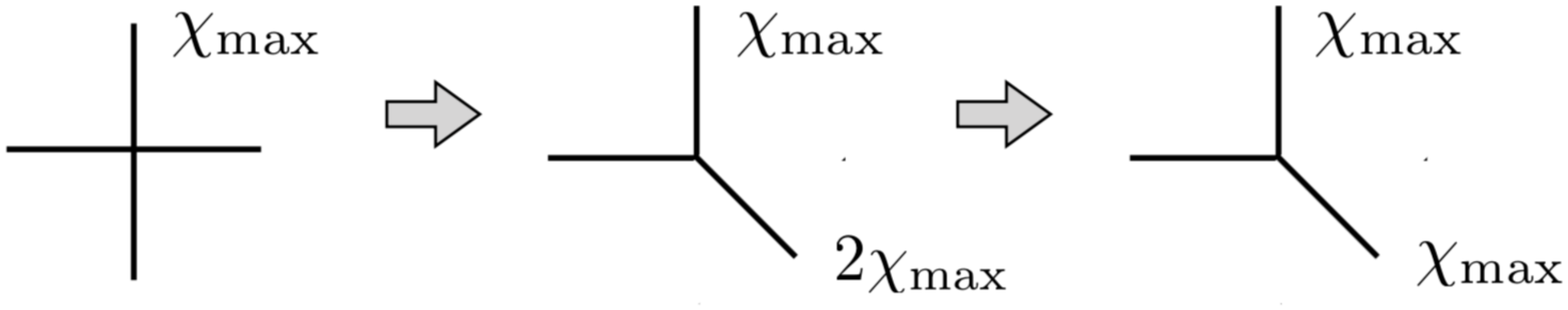}}} \hspace{-8mm} \ .
\label{tru}
\end{equation}
\vspace*{-2.4cm}

\noindent 
We will be interested in the cubic weights {\it before} implementing the truncation represented in second step.
At first order in the coupling constant, $V_n$ can be written as
\vspace*{-3.4cm}
\begin{eqnarray}
&& \vcenter{\hbox{\hspace{-22mm} \includegraphics[width=6.0cm]{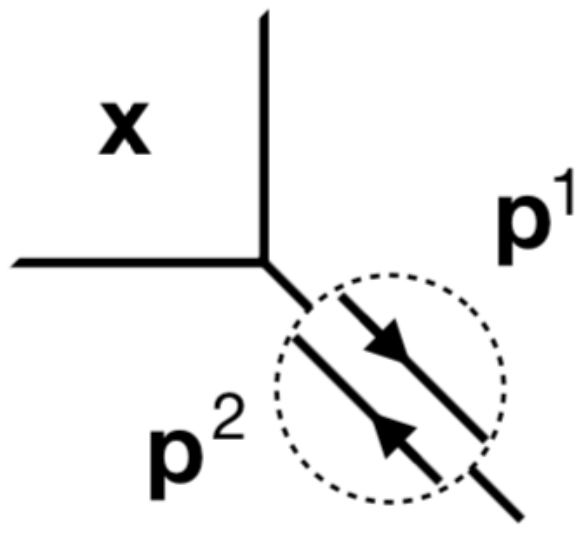}}} \hspace{-18mm}
 = \;\;\;e^{-{1 \over 2} {\bar {\bf x}}\big(A_n -{1\over 2} \id_2 \otimes D_{n-\!1}^{-1} \big){\bar {\bf x}} \,+\, i \,{\bar {\bf x}}\, U_n \, {\bar {\bf p}}}  \, \times \label{cubicone} \\[-38mm]
&& \hspace{18mm}\; \left[ {\hat g}_n \big({\partial}_{{\bf p}^1}, {\bar {\bf x}},{\partial}_{{\bf u}} \!\big) \; e^{-{\bf u} ( \id_2 \otimes D_{n-\!1}^{-1})  \, {\bf u}\,-\,{\bf p}^1\!  D_n^{-1}  
\, {\bf p}^1}  \right]_{{\bf u}={\bf x}^1} \ . \nonumber
\end{eqnarray}
\vspace*{-1mm}

\noindent The $\chi_{\rm max} \!\times \!\chi_{\rm max}$ matrix $D_{n-\!1}$ and the $2\chi_{\rm max} \!\times \!2\chi_{\rm max}$ matrices $A_n$, $U_n$ and $D_n$ are precisely the same as in the free case. 
Upon evaluating the derivatives $\partial_{\bf u}$, the differential operator ${\hat g}_n$ will transform into the level $n$ generalisation of \eqref{f1}.

For a $\lambda \phi^4$ theory,  the operator ${\hat g}_n$ 
can be represented in the terms of the quartic polynomial
\be
c_n \left[ 1+\lambda \Big( T^{(0)}_n + \sum_{i,j} T^{(2)}_{n,ij}\,  z_i z_j +   \sum_{i,j,k,l} \, T^{(4)}_{n,ijkl} \; z_i z_j z_l z_l  \Big)  \right]\ ,
\label{poly}
\ee
where the vector ${\bf z}$  stands for  $( \partial_{{\bf p}^1},{\bar {\bf x}}, \partial_{{\bf u}})$. 
The leading order constant $c_n$ 
has been factored out in the definition of the order-$k$ tensors ${T_n}^{(k)}$. Using these tensors we define
\be
\omega_{n,i}^{(2)}={1 \over \chi_{\rm max}} \, \sum_j |T^{(2)}_{n,ij}| \ , \hspace{3mm} \omega_{n,i}^{(4)}= {1 \over \chi_{\rm max}^3}\, \sum_{jkl} |T^{(4)}_{n,ijkl} | \ ,
\label{coeffs}
\ee
with $i=1,..,2\chi_{\rm max}$ running over the components of $\partial_{{\bf p}^1}$. 
These quantities provide an estimate of the relevance of $p^1_i$ in transmitting the effect of interaction. 
The larger the bond dimension, the smaller are some of the singular values in the diagonal matrices $D_n$.
If the derivatives $\partial_{\bf u}$ and $\partial_{{\bf p}^1}$ in \eqref{cubicone} were evaluated, small singular values would induce artificially large entries in the previous tensors. 
Such large contributions just mean that the probability distribution of the associated fields is close to a delta function
and, as discussed above, the TRG protocol can deal with these objects.

The singular values of $B_n$ and the above defined $\omega$'s are plotted in Fig.4a for $m=0.1$ at the coarse graining level $n=8$. 
We observe that the singular values follow a marked descending pattern.  On the contrary the $\omega$'s start slightly increasing and reach then a plateau extending up to $i \simeq 10$ . 
This implies that interaction strengthens the first set of entanglement ties, tending to make them equally relevant. However, although the entanglement hierarchy is modified, 
it is not incompatible with the leading order truncation criterium based on $B_n$. Indeed, 
the entries $\omega_{n,i}$ with $i > 14$ present a clear decreasing trend. The associated 
splitting fields  can thus be discarded with a small error both at the level of the leading and first order terms. This is in agreement with the good performance of $\chi_{\rm max}=16$ for $m=0.1$ seen in Fig.2.
It is interesting to notice the similarity between the $\omega$'s related to the two and four leg tensors.

\begin{figure}[h]
\begin{center}
\includegraphics[width=3.9cm]{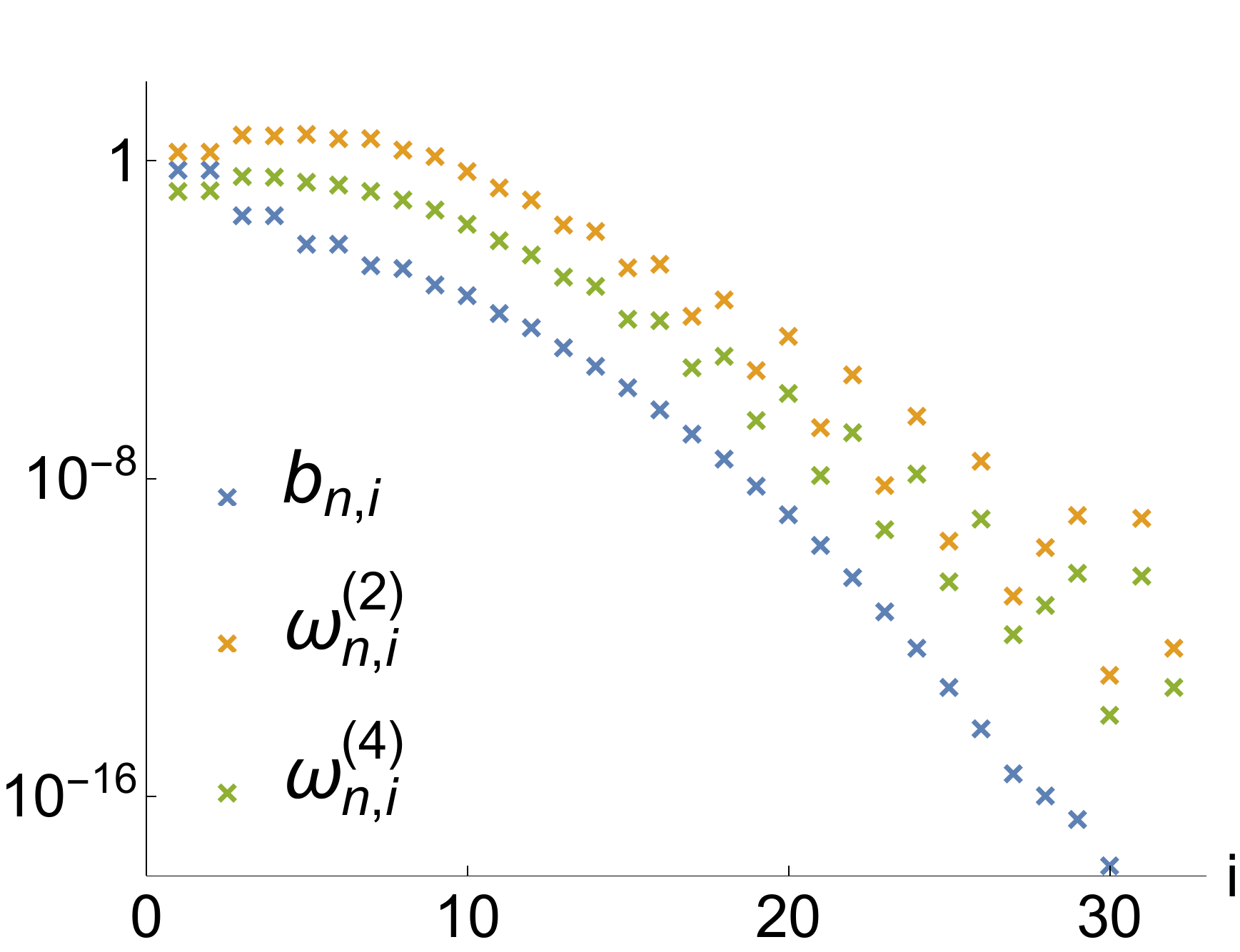}~~\includegraphics[width=3.9cm]{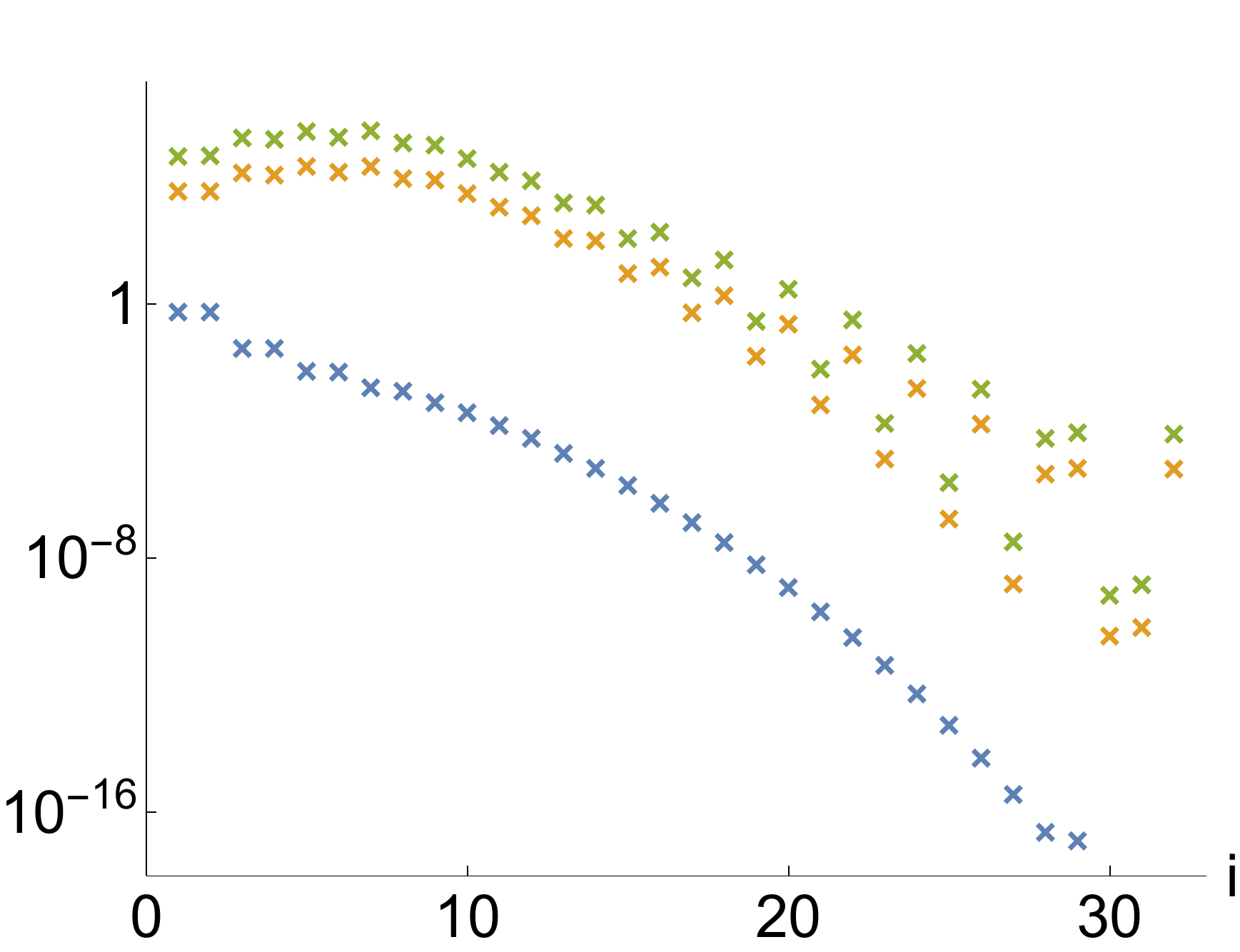}
\end{center}
\vspace*{-5mm}
\caption{\label{fig:fe2} Singular values of $B_n$ and sum of tensor elements \eqref{coeffs} for a) $m=0.1$ and b) $m=0.01$ at the coarse graining level $n=8$. The number of $b$'s and $\omega$'s at that level is $2 \chi_n=32$.}
\end{figure}

The behaviour shown in Fig.4a is general. Moreover, the smaller the mass the more $\omega$'s should be taken into account to obtain a good precision. This is observed in Fig.4b, which
suggests that including up to $i \simeq 20$ is necessary to reasonably describe the first order effects when $m=0.01$.
The pattern of the $\omega$'s also explains the two regimes seen in Fig.3b, exponential and power law, for the dependence of the relative
error $\delta f_1$ on the bond dimension. Only when $\chi_{\rm max}$ is large enough to include all fields associated with similarly relevant values of $\omega$, the convergence changes from exponential to power law.
On the other hand, the singular values of $B_n$ for $m=0.1$ and $0.01$ practically coincide. This agrees with the almost equal leading order error $\delta f_0$ seen in Fig.2 for both masses.

\subsection{CDL structure}

The standard TRG has the drawback of not being fully able to eliminate short range entanglement \cite{GW09}. The infrared (IR) fixed point of gapped systems treated with this protocol is given by a so-called corner double line structure (CDL)
\vspace*{-1.6cm}
\begin{equation}
\vcenter{\hbox{\includegraphics[width=3.6cm]{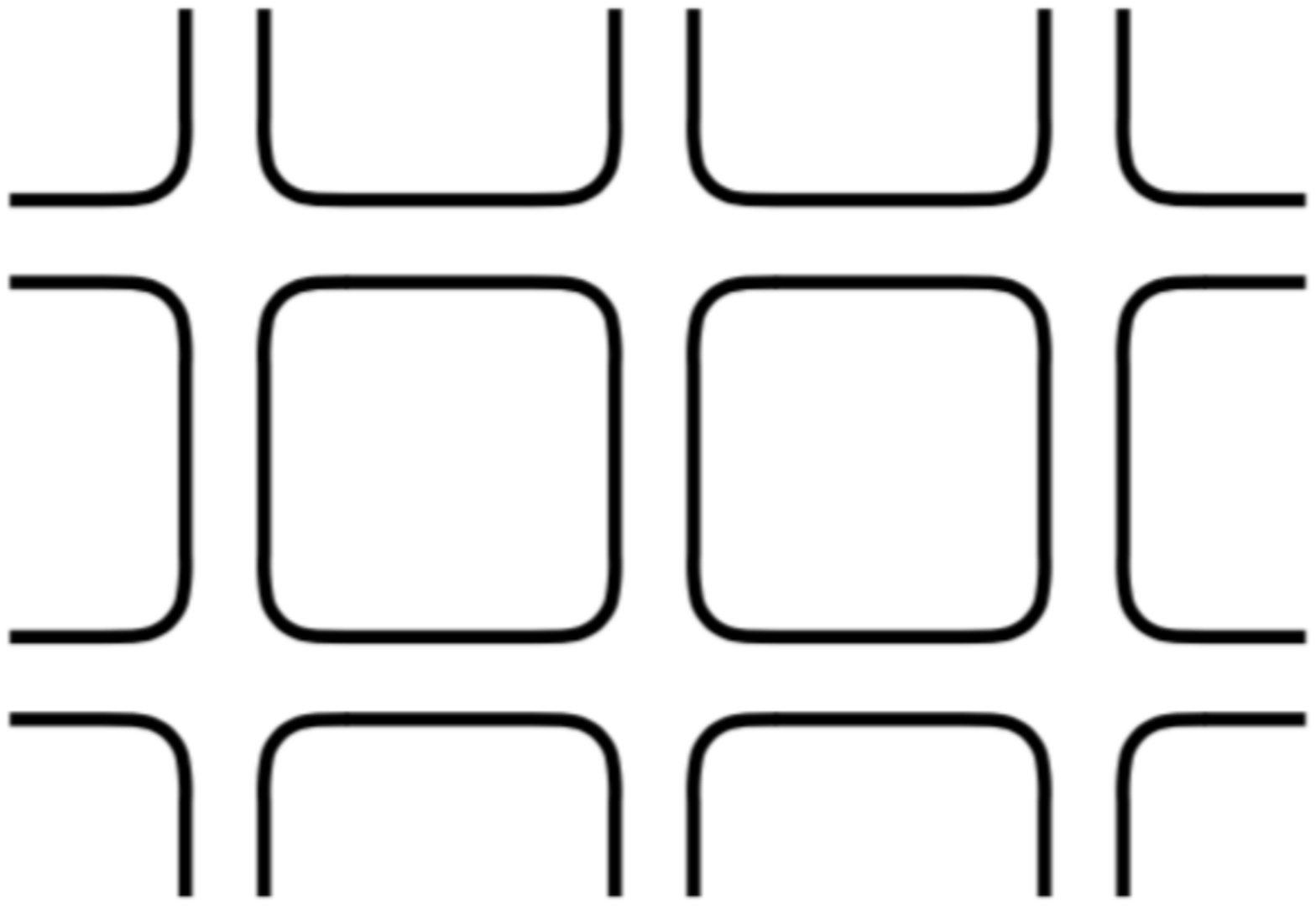}}} 
\label{CDL2}
\end{equation}

\vspace*{-1.6cm}
\noindent Correlations in the CDL network are confined inside each plaquette. A SVD based coarse graining fails however to detect this structure and promotes half of the CDL loops to the next level. 
This problem is replicated by the free boson TRG \cite{CS19}. Indeed the RG flow of the gaussian weights \eqref{Wgenfree} can be described by the evolution
\be
M_n \to M_{CDL} \ ,
\label{MCDL}
\ee
where the only non-zero entries of the matrix $M_{CDL}$ are those described by the CDL vertex.

When the degrees of freedom running in each plaquette decouple from the others, it is simple to integrate them out. In this sense the CDL is nothing but a redundant way to represent a trivial IR fixed point. This is confirmed by the ability of the TRG to transform the free boson lattice into a CDL network as the coarse graining process surpasses the scale set by the boson mass \cite{CS19}.
Having into account that the lattice spacing doubles every two TRG iterations, this happens when
\be
n > 2 \log_2 m^{-1} \ .
\label{IRscale}
\ee
We show in Fig.5a the RG evolution of the singular values of $B_n$ for $\chi_{\rm max}=16$ and $m=0.01$.  
At $n \simeq 18$ half of the singular values 
start to strongly decay. 
This signals the emergence of a CDL structure, since contributions from fields at opposite corners trivially factorize.
At the same time the CDL matrix \eqref{MCDL} freezes out, which is also observed in that figure.
The value $n=18$ is in good agreement with the theoretical estimation \eqref{IRscale}, whose {\it rhs} for $m=0.01$ gives $13.3$.
Notice that the quotient between the correlation length, $\xi \simeq m^{-1}$, and the lattice spacing at $n=18$ is $ \simeq 0.2$.

We are ready now to analyze the RG flow associated with the first order correction to the free energy. 
The tensors defined in \eqref{poly} encode the structure at first order in the perturbative expansion of the $\lambda \phi^4$ theory.
They should reach an IR fixed point at the same scale as the 
singular values $b_{n,i}$. 
A modification of the IR scale \eqref{IRscale} can only arise from an all order resumation of perturbative effects, as that implied by the usual RG equations.
We are not addressing this issue here. 
Consistently, we found that the two and four leg tensors freeze out right after the leading gaussian structure reaches the CDL fixed point. 

\vspace*{-2mm}
\begin{figure}[h]
\begin{center}
\includegraphics[width=4cm]{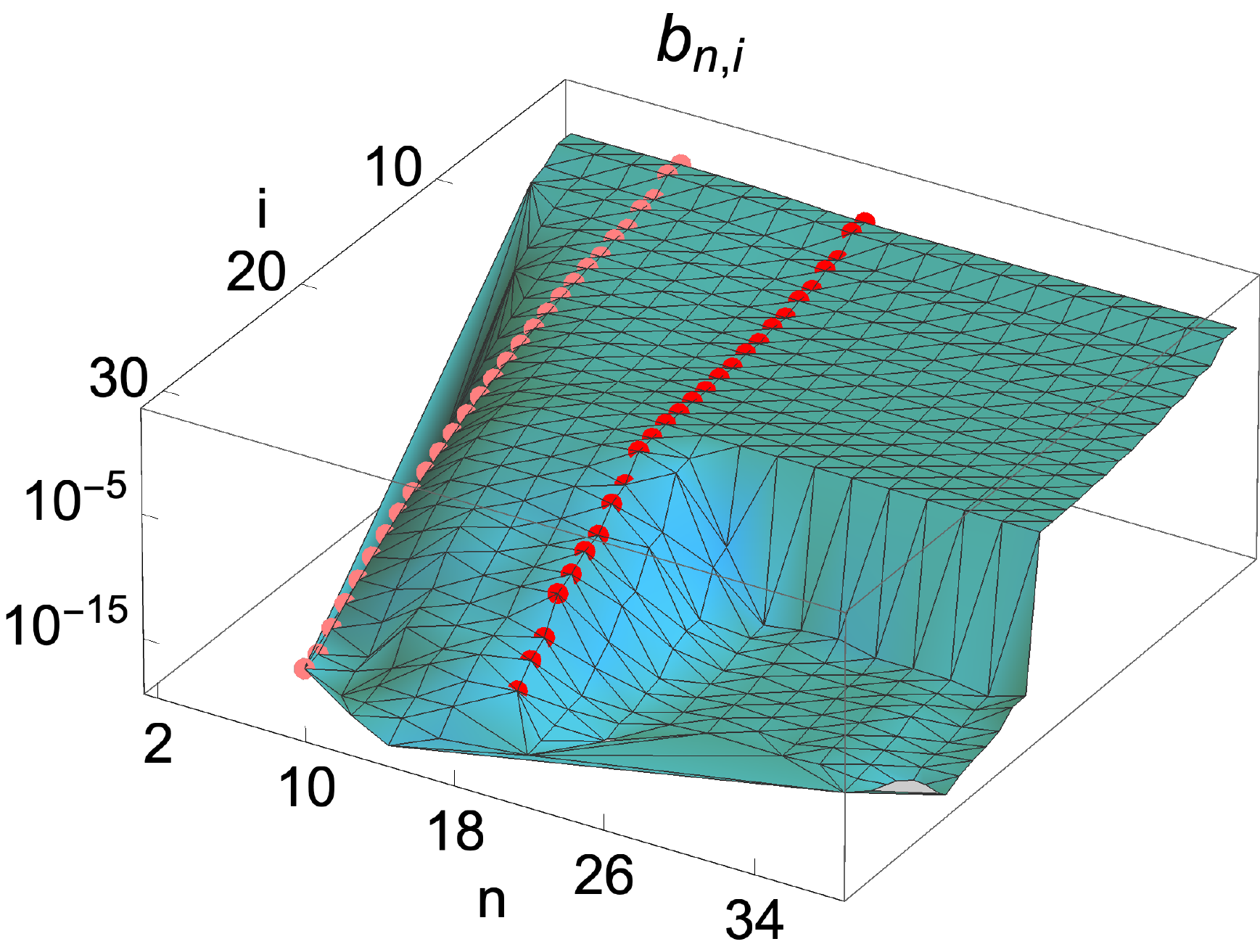}\includegraphics[width=4cm]{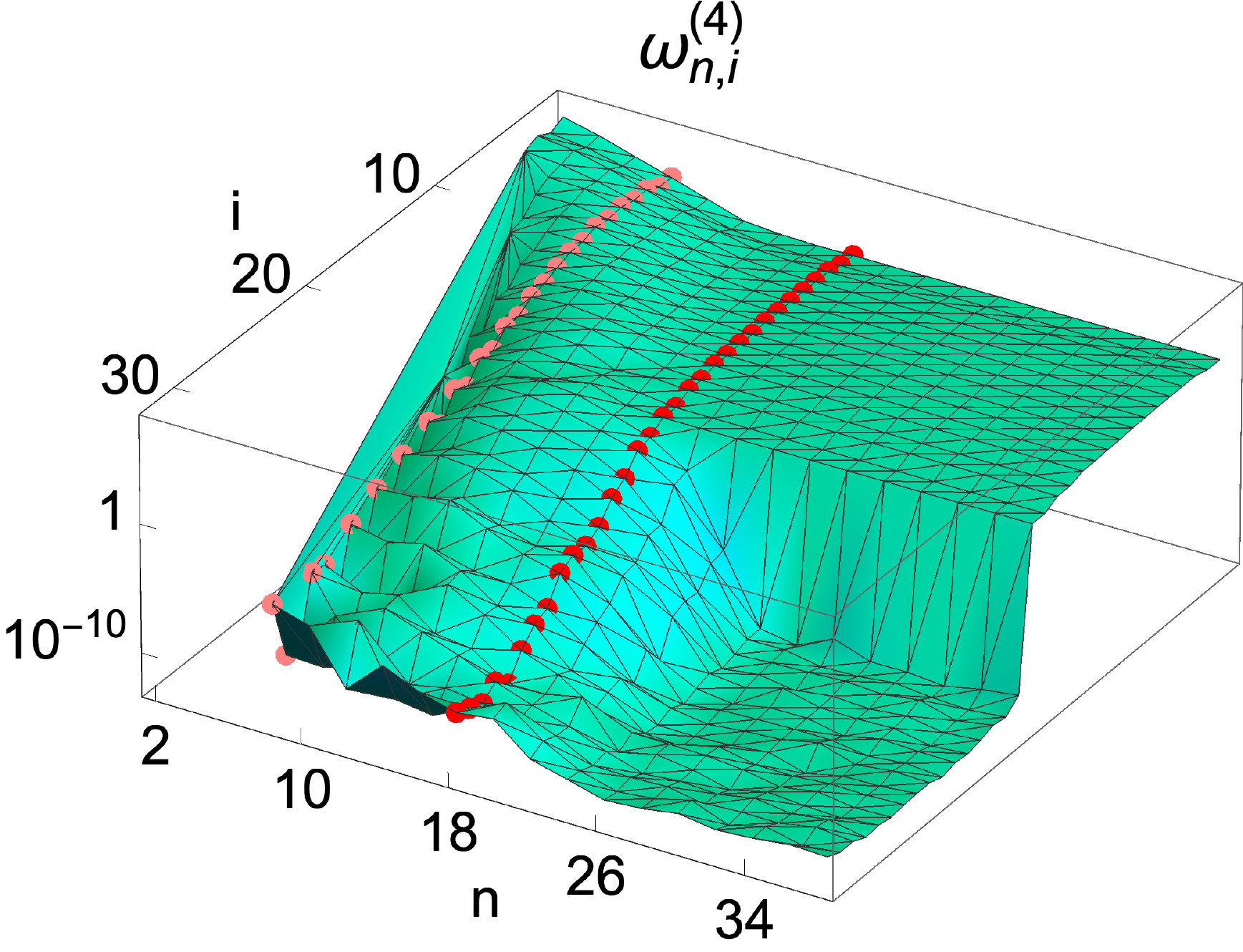}
\end{center}
\vspace*{-5mm}
\caption{\label{fig:fe3} RG flow of a) the singular values of $B_n$ and b) the sum of ${T_n}^{(4)}$ elements \eqref{coeffs} for $\chi_{\rm max}=16$ and $m=0.01$. The pink line signals $n=8$, the iteration where truncation begins with $\chi_{\rm max}=16$, and the red line marks $n=18$. These plots refer $n$ even, horizontal lattices. Analogous results describe $n$ odd, tilted lattices.}
\end{figure}

The RG flow of the order $\lambda$ terms can be portrayed with the help of 
\eqref{coeffs}. We have plotted in Fig.5b the evolution of the $\omega$'s associated with ${T_n}^{(4)}$
for $\chi_{\rm max}=16$ and $m=0.01$. Not only these quantities stabilise at the same time as $b_{n,i}$, 
but together with them, the corresponding half becomes zero. The same result holds for ${T_n}^{(2)}$.
This implies that the associated ${\bf p}$ in \eqref{cubicone} are irrelevant to the factorization of both gaussian and interacting cross terms. 
Hence the truncation in the second step of \eqref{tru} is automatic, which is a distinct characteristic of the CDL structure.

Finally, we have checked that the 
two and four leg tensors reproduce the CDL pattern of network connections.  
To that aim we define the matrix
\be
\Omega_{ij}= |T_{n,ij}^{(2)}|\,+\,{1 \over \chi_{\rm max}^2} \, \sum_{k,l} | T^{(4)}_{n,ijkl} | \ .
\label{omat}
\ee 
Recall that its indices refer to $( \partial_{{\bf p}^1},{\bar {\bf x}}, \partial_{{\bf u}})$. Contrary to \eqref{coeffs}, we assume that the truncation in the second step of \eqref{tru} has been already implemented and
the bond dimension in all links is $\chi_{\max}$.
A non-zero entry $\Omega_{ij}$ reflects that the Feynman diagrams connect components $i$ and $j$.
A graphical representation of this matrix is given in Fig.6 for $\chi_{\rm max}=16$, $m=0.01$ and $n=24$, when the CDL regime is clearly stablished for the leading gaussian terms.
The boxes defined by the thin black lines have dimension $\chi_{\rm max}\!\times \!\chi_{\rm max}$, while the numbers on their side label the links of the CDL cubic weight included in that figure.
The colour code describes the magnitude of the entries, making clear the CDL structure. 
We obtain thus a unified picture of the RG flow at leading and first order in the coupling constant.

\vspace*{-17mm}
\begin{figure}[h]
\begin{center}
\includegraphics[width=10cm]{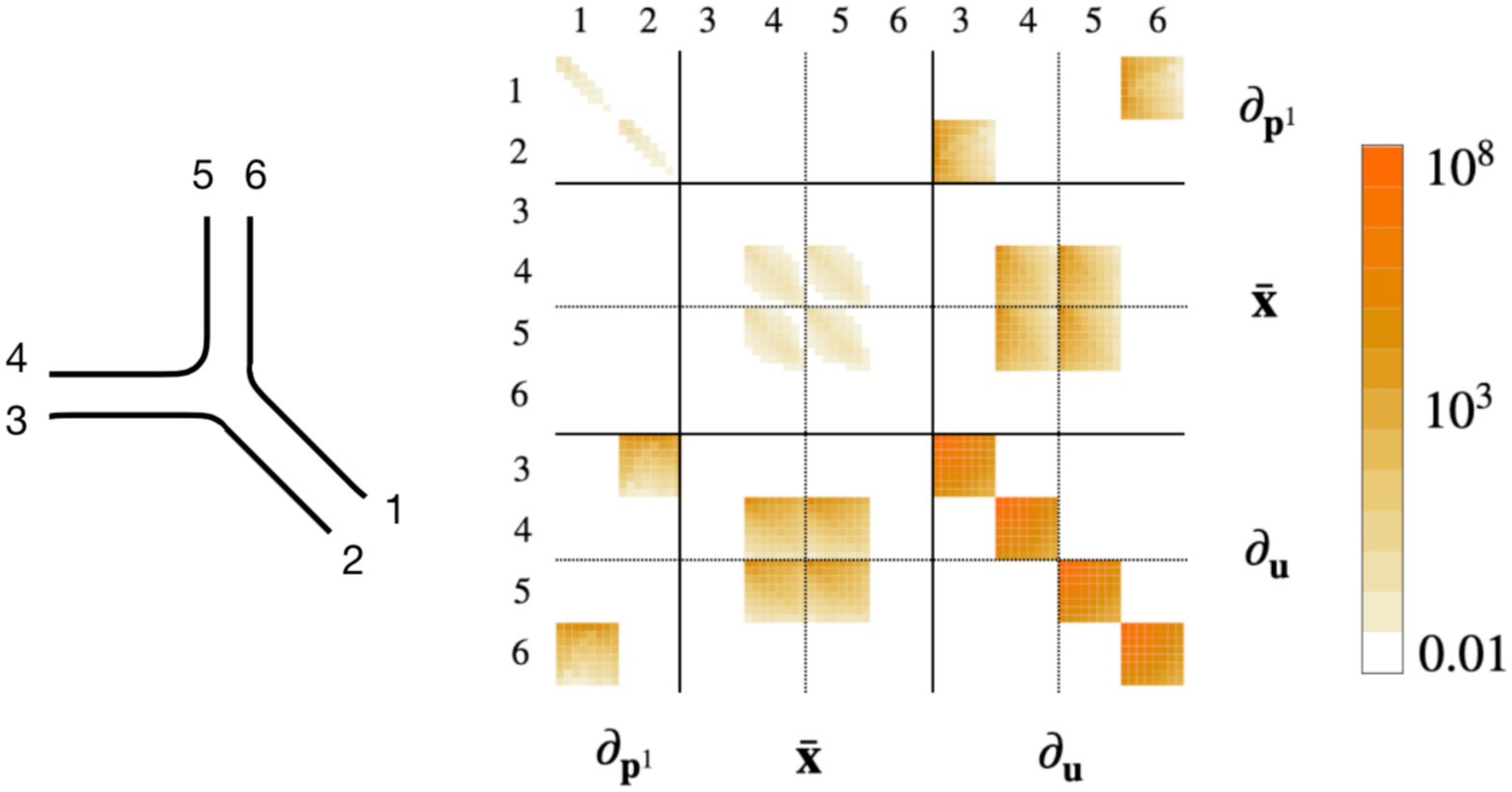} \hspace{10mm}
\end{center}
\vspace*{-22mm}
\caption{\label{fig:fe4} Colour plot of the matrix \eqref{omat} for $\chi_{\rm max}=16$, $m=0.01$ and $n=24$.}
\end{figure}
\vspace*{-5mm}

\section{Conclusions}

We have explored the formulation of a real space coarse graining protocol for interacting QFT's based on local manipulations on continuous variables.
Together with an entanglement rooted organisation of degrees of freedom, we have used the standard Wilsonian approach as guideline.
This led to move the SVD analysis characteristic of real space RG algorithms from the Boltzmann weights, which describe the discretized QFT partition function, to their exponent or in other words, to the free energy. Our motivation was to operate directly at the level of fields. Each field is a proxy for an infinite subset of the singular values that would result from a standard SVD.
The strong structure that this automatically implies aims to be an efficient theoretical tool, able to obtain insights that are difficult for more numerical oriented algorithms.

We have developed a framework based on the TRG network \cite{LN07}, which realises this program in the context of perturbation theory. 
Describing the effect of interaction in terms of local manipulations requires the introduction of extra fields with distinct properties. They have a trivial leading order structure and their number increases 
with the order in perturbation theory at which we want to work.
Due to the simple nature of the associated quantum corrections,  the role of these additional fields at first order in the coupling constant is secondary.
It would be therefore important to extend the detailed numerical analysis we have performed here beyond first order.

The adapted TRG network evaluates the partition function of a free boson with excellent precision and a numerical cost independent of the boson mass  \cite{CS19}. 
We have chosen a $\lambda \phi^4$ theory to benchmark the performance at first order in perturbation theory. 
In this case the numerical cost increases at small masses, but the results exhibit a fast convergence with the bond dimension.
Using the direct correspondence between the field content of the free and first order networks, we have analyzed how the interaction modifies the leading order entanglement hierarchy.
The quantum corrections work towards levelling off the pattern of marked decreasing relevance that organizes the fields in the absence of interaction.
This effect involves a larger subset of fields the lower the mass is, explaining a dependence of the numerical cost on the mass which was not present in the free network.
Besides, we obtained a consistent picture of the RG flow at leading and first order. As it is characteristic of standard TRG algorithms, a fixed point with CDL structure was attained in the IR.

At higher order in perturbation theory, our algorithm would have to deal with tensors of growing rank and dimension. At order $N$, and assuming we can deal only with connected Feynman diagrams, the analogue of expansion \eqref{poly} will contain tensors up to rank $2N+2$. 
The dimension of each tensor index is $2^{N} 5 \chi_{\text{max}}$, where we have used the results from Section V.C.
Following Section VI, $\chi_{\text{max}}$ refers to the bond dimension of the free case.
We expect that the special characteristics of the extra fields \eqref{newfields} may help to simplify the algorithm.

Tensor networks are designed to deal with any value of the coupling constant. The perturbative approach we have followed here might seem thus limited.
We believe however that it is an important step towards a goal that is extremely difficult for QFT. 
Although beyond the scope of this paper, a venue to include non-perturbative effects could be adding finite $\lambda$ corrections to the gaussian structure 
that underlies the algorithm.
A modification of the leading order gaussian structure can also be relevant for improving the performance in the limit of small masses.
On the other hand, we would like to stress that the perturbative framework is not essential for the SVD protocol presented in Section IV.B.

Finally, there are other interesting and simpler generalizations of the adapted TRG protocol. In particular, the calculation of correlation
functions and its application to field theories with fermions. We plan to address some of these issues in the near future.

\begin{acknowledgments}

We thank Pasquale Calabrese, Jos\'e Melgarejo, Javier Molina-Villaplana, Giuseppe Mussardo, Antoine Tilloy  and Erik Tonni for conversations.
The work of M.C. is supported by a contract BES-2017-080586.
We acknowledge  financial support from the grants 
PGC2018-095862-B-C21, QUITEMAD+ S2013/ICE-2801 and SEV-2016-0597 of the ``Centro de Excelencia Severo Ochoa'' Programme.

\end{acknowledgments}

\onecolumngrid

\appendix

\section{SVD vs gaussian SVD}

In this appendix we compare the standard SVD with the gaussian SVD employed throughout the paper
as the basic mathematical  tool to implement the gTRG. The SVD is equivalent to the Schmidt
decomposition of a quantum state. We shall use this correspondence in what follows.

\subsection{The Fourier transform as a continuous SVD} 

Let us start with eq.\eqref{fourier}
 
\beq
e^{ - \frac{1}{2} (x_1 - x_2)^2} = \int \frac{dp}{ \sqrt{2 \pi}} e^{ i p( x_1 - x_2) - \frac{1}{2} p^2}  \ , 
\label{1}
\eeq
and define the state
\beq
| \psi \rangle = \int dx_1 dx_2 \, e^{ - \frac{1}{2} (x_1 - x_2)^2}  |x_1 \rangle \otimes |x_2 \rangle \, , 
\label{2}
\eeq
where $|x \rangle$ is a orthonormal basis in the Dirac sense, that is 
$\langle x |x' \rangle = \delta(x - x')$.  Eq.\eqref{2} represents a pair of particles
in a bound state that using \eqref{1} can be written as 
\barray 
| \psi \rangle & =  &    \sqrt{2 \pi} \int dp  \;  e^{ - \frac{1}{2} p^2}  |p \rangle \otimes |-p \rangle \ , 
\label{5}
\earray
where the states 
\beq
| p \rangle \equiv 
\int \frac{dx}{\sqrt{2 \pi}}  \, e^{ i p x}  |x \rangle \ , 
\label{6}
\eeq
satisfy
\beq
\langle p | p' \rangle = \delta( p - p') \ . 
\label{7}
\eeq
Eq.\eqref{5} is the Schmidt decomposition of the state \eqref{2} in  the 
orthonormal basis $| p \rangle$. Eq. \eqref{1} is a continuous SVD,  that is nothing but the Fourier transform 
of $e^{ - \frac{1}{2} p^2}$.  Notice that the norm of  \eqref{5} is $\langle \psi |\psi \rangle = 2 \pi^{3/2} \delta(0)$
that reflects the translational invariance of the two body wave function $e^{ - \frac{1}{2} (x_1 - x_2)^2}$.

\subsection{SVD of a normalized gaussian state}

Let us next consider the function 

\beq
W(x_1, x_2) = \rho  e^{ - \frac{a}{2} (x_1^2 + x_2^2) + b x_1 x_2} \ , 
\label{29}
\eeq
that has the same form as eq.(12).  We define the state 
\beq
| W \rangle = \int dx_1  dx_2  \;  W(x_1,x_2) |x_1 \rangle \otimes | x_2 \rangle \ , 
\label{36}
\eeq
that is normalized provided $a > |b|$,  with the normalization factor $\rho =  \pi^{- 1/2} (a^2 - b^2)^{ 1/4}$. 
The state \eqref{2} corresponds to $a=b=1$ that, as explained above, is normalized in the Dirac sense. 

To find the Schmidt decomposition of \eqref{36} we use the identity  \cite{A70} 
\beq
\sum_{n=0}^\infty \frac{u^n }{2^n n!} H_n(x) H_n(y)  = \frac{ 1}{ \sqrt{1- u^2}}  {\rm exp} \left(  \frac{ 2 u}{ 1 + u} x y  - 
\frac{u^2}{ 1 - u^2} ( x - y)^2 \right)  \ , 
\label{30}
\eeq
where $H_n(x)$ are the Hermite polynomials that give the eigenfunctions of the
Hamiltonian of the  harmonic oscillator $H = \frac{1}{2} ( - \partial_x^2 + x^2)$, 
\beq
\psi_n(x) = \frac{1}{ \sqrt{ \pi^{1/2} 2^n n!}} H_n(x) e^{ - x^2/2}  \ . 
\label{31}
\eeq
Using \eqref{31} one can write \eqref{30} as

\beq
\sum_{n=0}^\infty u^n  \psi_n(x) \psi_n(y)  = \frac{ 1}{   \sqrt{\pi(1- u^2)}}  {\rm exp} \left( - \frac{ x^2 + y^2 - 2 u x y}{ 2 (1 - u^2)}  \right) \ . 
\label{32}
\eeq
Comparing \eqref{29} and \eqref{32} we find
\beq
a x_1^2 = \frac{ x^2}{ 1- u^2}, \qquad a x_2^2  = \frac{y^2}{ 1 - u^2}, \qquad b x_1 x_2 = \frac{ u xy}{ 1 - u^2} \ , 
\label{33}
\eeq
so that 
\beq
u^2 =  \left( \frac{b}{a} \right)^2  \ . 
\label{34}
\eeq
Let us write  \eqref{29} using the variables $x$ and $y$ 
\beq
W(x, y) = \frac{ 1}{   \sqrt{\pi}}  {\rm exp} \left( - \frac{ x^2 + y^2 - 2 u x y}{ 2 (1 - u^2)}  \right) , \quad |u | < 1 \, 
\label{35}
\eeq
where we took into account the Jacobian of the transformation. The state \eqref{36} becomes
\beq
| W \rangle = \sqrt{1 - u^2} \sum_{n=0}^\infty   u^n   |n \rangle \otimes | n \rangle \, , 
\label{37}
\eeq
where 
\beq
|n \rangle = \int dx \, \psi_n(x) |x \rangle \ . 
\label{38}
\eeq
Equation  \eqref{37} is  the Schmidt decomposition
of $|W\rangle$  in terms of  the discrete Schmidt coefficients, 

\beq
w_n = \sqrt{1 - u^2}  \; u^n,  \quad u = 0, 1, \dots, \infty \ . 
\label{39} 
\eeq
Talking the trace over the left, or right, Hilbert space of the pure density matrix of \eqref{37} gives
\beq
\rho = {\rm tr}_{\rm L} | W \rangle \langle W| = (1 - u^2) \sum_{n=0}^\infty u^{2 n} | n \rangle \langle n| \, 
\label{39b}
\eeq
whose  von Neumann entropy is 
\beq
S_1 = - \log (1 - u^2) - \frac{ u^2 \log u^2}{ 1 - u^2} \ , 
\label{41}
\eeq
that diverges in the limit $u^2 \rightarrow 1^-$.

\subsection{Gaussian SVD of a normalized gaussian state}

Let us consider again the state \eqref{36}, and write  \eqref{29} as 
\beq
W(x_1, x_2) = \rho e^{ - \frac{1}{2} (a -b) (x_1^2 + x_2^2) - \frac{b}{2} (x_1 - x_2)^2} \ . 
\label{29b}
\eeq
Using \eqref{1} with $x_{1,2} \rightarrow \sqrt{b} x_{1,2}$, this equation  reads
\beq
W(x_1, x_2) = \rho e^{ - \frac{1}{2} (a -b) (x_1^2 + x_2^2)}   \int \frac{dp}{ \sqrt{2 \pi b}} e^{ i p( x_1 - x_2) - \frac{1}{2b} p^2} \ ,
\label{29c}
\eeq
that allow us to write \eqref{36} as 
\beq
| W \rangle = \frac{ \rho}{ \sqrt{ 2 \pi b} } \int dp \, e^{ - \frac{1}{2 b} p^2} |p, a-b \rangle \otimes |- p, a-b \rangle \ , 
\label{29d}
\eeq
where 
\beq 
| p, a-b \rangle = \int dx \,  e^{ - \frac{1}{2} ( a- b)  x^2  + i p x}  | x \rangle  \ .
\label{29e}
\eeq
These states are not orthogonal since
\beq
\langle p, a-b | p' , a-b \rangle = \sqrt{ \frac{ \pi }{a-b} } e^{ - \frac{ (p - p')^2}{ 4 (a-b)}} \ . 
\label{29f}
\eeq
Equation \eqref{29d} looks like  a continuous Schmidt decomposition  similar to \eqref{5}, 
but it is not because
the basis $|p, a-b\rangle$ is not orthogonal when $a > |b|$. On the other hand,  we saw above that the
Schmidt decomposition of $| W \rangle$ involves an infinite number of terms but is  discrete (see \eqref{37}).
The gaussian SVD is an example of a look alike Schmidt decomposition that is forced upon  us 
by the condition of working with the exponent of the Boltzmann weights or the effective action
in Quantum Field Theory.

\section{Loop matrices}

We will construct the matrices $Q_n$ and $C_n$ that govern the gaussian kernel of the TRG coarse graining integrals. 
The leading gaussian terms do not couple the 
fields with a counterpart in the free model, the set \eqref{chain}, with the new fields \eqref{newfields} introduced by interaction. 
For the $n=2$ matrices, this was made explicit in \eqref{M22}.
Based on it, we will focus on the matrix blocks associated with the set \eqref{chain}.
They are given by the same expressions that apply to the free network \cite{}. 
In order to simplify notation, we refer to the free case in the following presentation.

In an appropriate basis, and up to a propotionality constant, 
the Boltzmann weights of the free model are described by
\vspace*{-3.5cm}
\be
 \vcenter{\hbox{\hspace{-32mm} \includegraphics[width=6.1cm]{WW.pdf}}} \hspace{-1.9cm} = \;\;\;\;\; e^{-{1 \over 2} {\bf x} \, M_n{\bf x}} \ , \hspace{1.5cm} M_n=\begin{pmatrix} A_n & 0 \\
                              0 & A_n \end{pmatrix} +  \begin{pmatrix} B_n  & -B_n \\
                                     -B_n & B_n \end{pmatrix} \ .                               
\label{Mfree2}                                  
\ee  
\vspace*{-3.7cm}

\noindent The $2 \times 2$ block structure of $M_n$ corresponds to the choice ${\bf x}=({\bf x}_L,{\bf x}_R)$ with ${\bf x}_L=({\bf x}_1,{\bf x}_2)$ and ${\bf x}_R=({\bf x}_4,{\bf x}_3)$, motivated by
a subsequent splitting of $W_n$ along the axis $12-34$. The factorization of the Boltzmann weights is based on the SVD of $B_n$, which rewrites this matrix as $U_n D_n U_n^T$. In general $U_n$ is a $2\chi_n \times \chi_{n+1}$ isometry. We have $\chi_{n+1} \leq 2\chi_n$, since some singular values in $D_n$ might be truncated because their number is larger than $\chi_{\rm max}$ or discarded because they vanish. 
The matrices $A_n$ and $U_n$ have a further interesting structure \cite{CS19}
\be
A_n = {1 \over 2} \id_2 \otimes  D_{n-1}^{-1}\, +  \begin{pmatrix}a_n  & -a_n \\
                                    -a_n & a_n \end{pmatrix} \ , \hspace{1cm} 
U_n={1 \over \sqrt{2}} \begin{pmatrix} u_n & v_n \\
                              u_n & -v_n \end{pmatrix}        \ .                           
\ee
The matrix $a_n$ is real symmetric and positive. Both $a_n$ and $D_{n-1}$ have dimension $\chi_n \! \times \! \chi_n$. The isometries $u_n$ and $v_n$ have real entries and dimension $\chi_n \times {1 \over 2}\chi_{n+1}$, satisfying $u_n^+ u_n= v_n^+ v_n = \id$.

After splitting $W_n$ into two cubic weights according to the $2 \times 2$ block structure in \eqref{Mfree2}, four cubic weights are sewed together 
in order to obtain $W_{n+1}$.
The kernel of the coarse graining integral that eliminates the fields ${\bf x}$ in favour of the next level ${\bf p}$ is 
\vspace*{-1.8cm}
\be
 \vcenter{\hbox{\hspace{-10mm} \includegraphics[width=8.1cm]{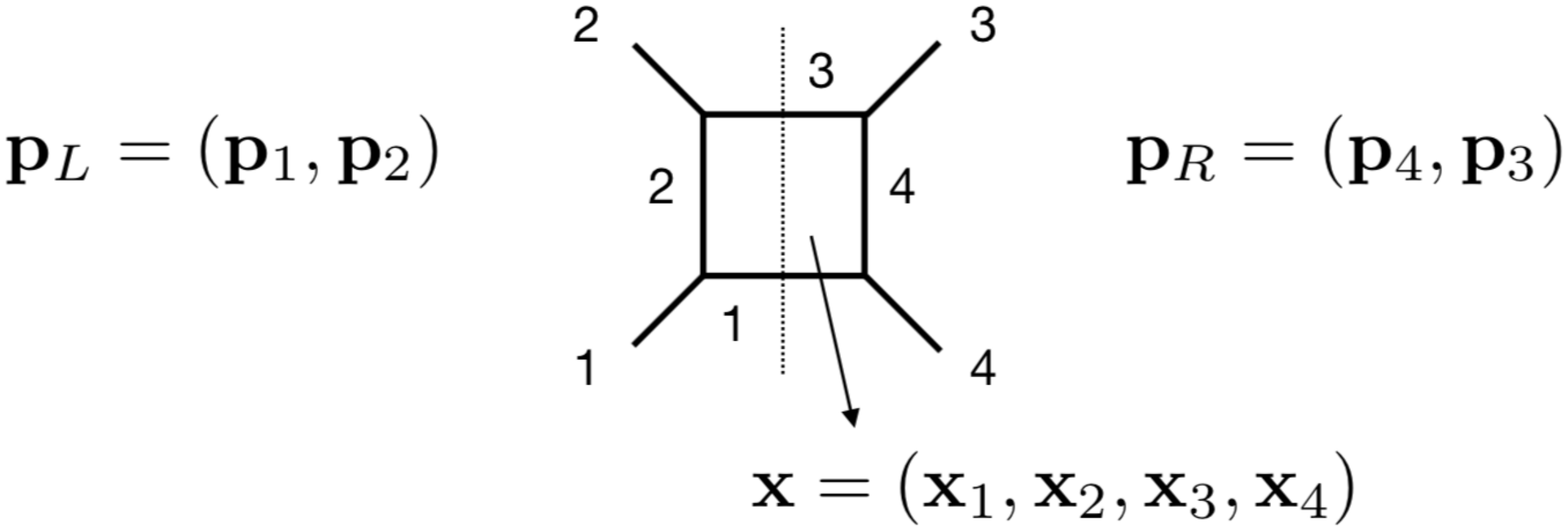}}} \hspace{0.0cm} = \;\;\;\;\; 
 e^{-{1 \over 2} {\bf x} \, Q_n{\bf x}+i {\bf x} \, C_n {\bf p}} \ ,
\ee
\vspace*{-1.7cm}

\noindent
where we have ignored again a multiplicative constant. The $4 \chi_n \times 4 \chi_n$ matrix $Q$ is readily built out of $D_{n-1}^{-1}$ and $a_n$
\be
Q_n= \id_4\otimes D^{-1}_{n-1} \, + \begin{pmatrix} 2 & -1 & 0 & -1\\
                                                                               -1 & 2 & -1 & 0 \\
                                                                               0 & -1 & 2 & -1 \\
                                                                                -1 & 0 & -1 & 2 \end{pmatrix} \otimes a_n \ .
\label{QQ}
\ee                              
It is convenient to decompose the matrix $C_n$ into two components such that $C_n {\bf p}=  C_{n,L} {\bf p}_L +C_{n,R} {\bf p}_R$. The $4 \chi_n \times 2 \chi_{n+1}$ matrices $C_{n,LR}$ are constructed in terms of the isometries $u_n$ and $v_n$ 
\be
C_L={1 \over \sqrt{2}} \begin{pmatrix} u_n & v_n & 0 & 0\\
                                  u_n  & -v_n & -u_n & v_n \\
                                   0 & 0 & -u_n & -v_n \\
                                    0 & 0 & 0 & 0 \end{pmatrix}  \ ,  \hspace{1.5cm}
C_R ={1 \over \sqrt{2}}\begin{pmatrix} -u_n & -v_n & 0 & 0\\
                                  0 & 0 & 0 & 0 \\
                                   0 & 0 & u_n & v_n \\
                                    -u_n & v_n & u_n & -v_n \end{pmatrix}  \ .
\label{CLR}
\ee
\noindent The zeroes in the last row of $C_{n,L}$ reflect that ${\bf x}_4$ does not couple to ${\bf p}_L$ and correspondigly for the second row of zeroes in $C_{n,R}$.
Generalizing \eqref{M2} and \eqref{M22}, we obtain
\be 
A_{n+1}={1 \over 2} \Big[ \id_2 \otimes  D_{n}^{-1}  + \, (C_{n,L}\!-C_{n,R})^T \, Q^{-1}_n (C_{n,L}\! -C_{n,R}) \Big]
\ , \hspace{1cm}  B_{n+1}= - C_{n,L}^T \, Q^{-1}_n C_{n,R} \ , 
\label{CQC}
\ee
where from \eqref{Mfree2}  we have used 
\be
C_{n,L}^T \, Q^{-1}_n C_{n,L} = C_{n,R}^T \, Q^{-1}_n C_{n,R} \ , \hspace{1.5cm} C_{n,L}^T \, Q^{-1}_n C_{n,R} =C_{n,R}^T \, Q^{-1}_n C_{n,L} \ .
\ee

\vspace*{5mm}

\section{Feynman rules}

We present the Feynman rules for building the diagrammatic expansion of $F_1$ \eqref{F1}
\be
F_1({\bf p})=\int d {\bf x}\;  e^{-{1 \over 2} {\bf x} \, Q_0 \, {\bf x}} \; \prod_{i=1}^4 \, f_{0}^{(i)}({\bf x}+i Q_0^{-1} C_0 {\bf p} ) \ .
\label{F1bis}
\ee
In analogy with the Wilsonian effective action reviewed in Section III, double lines are used to represent the fields integrated over (${\bf x} \equiv$ fast fields) and single lines for the next level fields 
(${\bf p} \equiv$ slow fields).
The propagator of the fast fields is given by
\vspace*{-2.5cm}
\begin{equation}
 \vcenter{\hbox{\hspace{-24mm} \includegraphics[width=3.9cm]{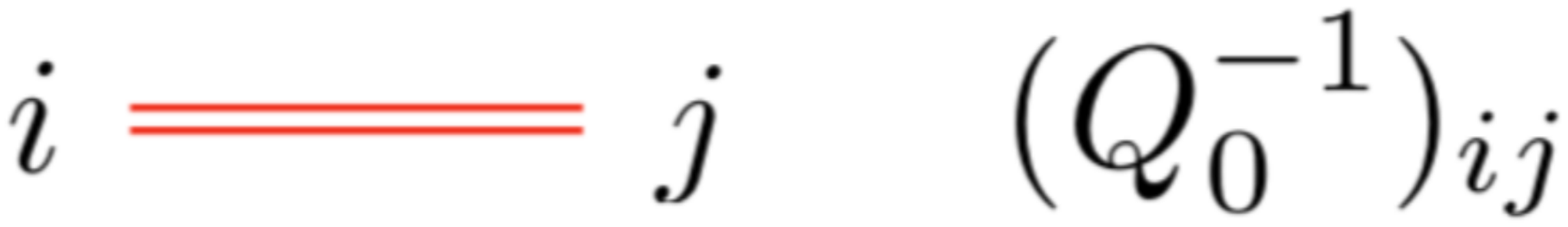}}} \hspace{-0mm} \ , \hspace{0mm}
\label{rules}
\end{equation}
\vspace*{-2.7cm}

\noindent with $i=1,..,4$. The kinetic terms and spatial derivatives in the original lagrangian \eqref{lagrangian} render the matrix $Q_0$ non-diagonal. As a consequence $Q_0^{-1}$ propagates linear combinations 
involving all four ${\bf x}$-fields in the integral \eqref{F1bis}. On the contrary, interaction in the initial lattice only couples fields living on the same link. 
Using a $\lambda \phi^4$ theory as example, the interaction vertex for the ${\bf x}$ fields is
\vspace*{-0.5cm}
\begin{equation}
 \vcenter{\hbox{\hspace{-20mm} \includegraphics[width=3.3cm]{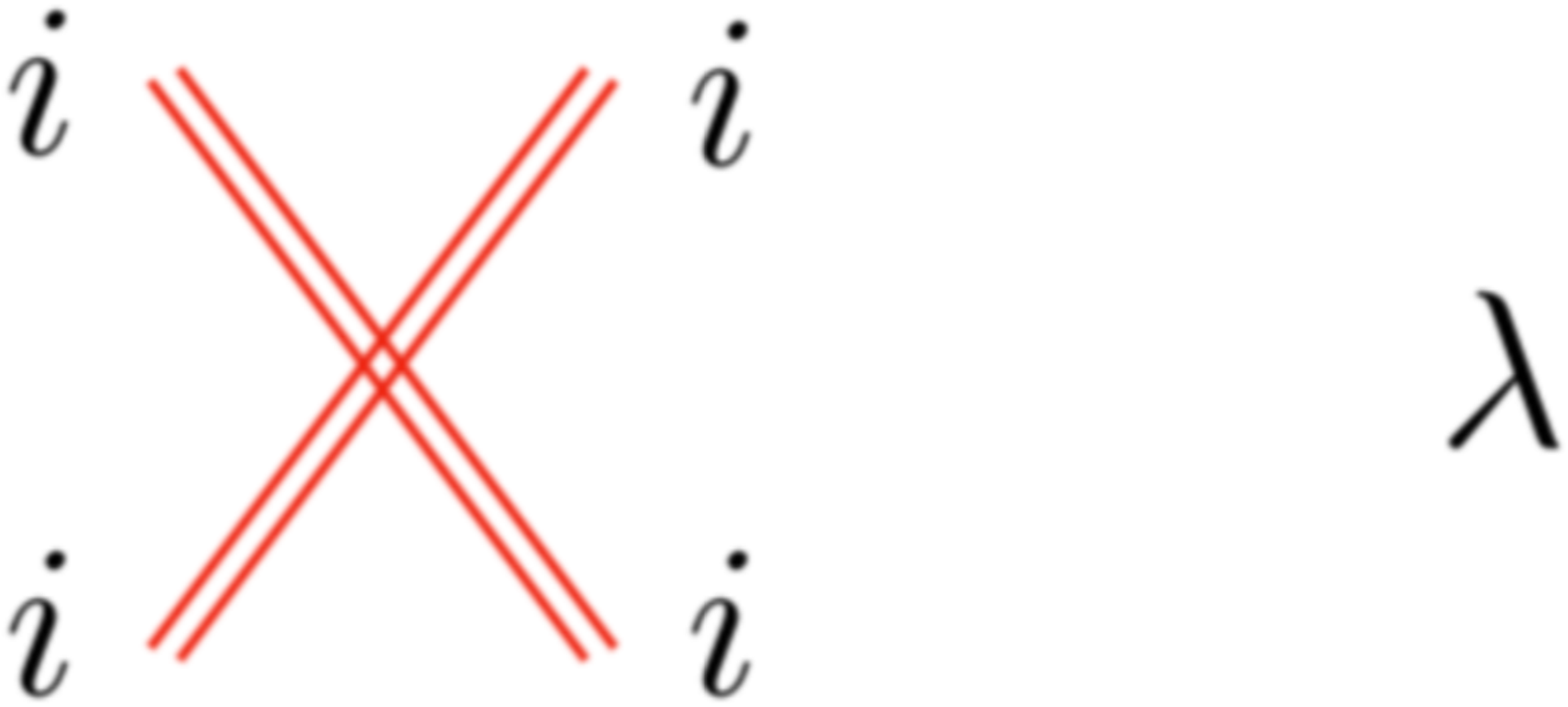}}} \hspace{-0mm} \ . \hspace{8mm}
\label{rules2}
\end{equation}
\vspace*{-0.6cm}

\noindent The interaction vertices between fast and slow fields are obtained by dressing some double lines of the vertex \eqref{rules2} with $Q_0^{-1} C_0$
\vspace*{-3.4cm}
\begin{equation}
 \vcenter{\hbox{\hspace{-12mm} \includegraphics[width=11.5cm]{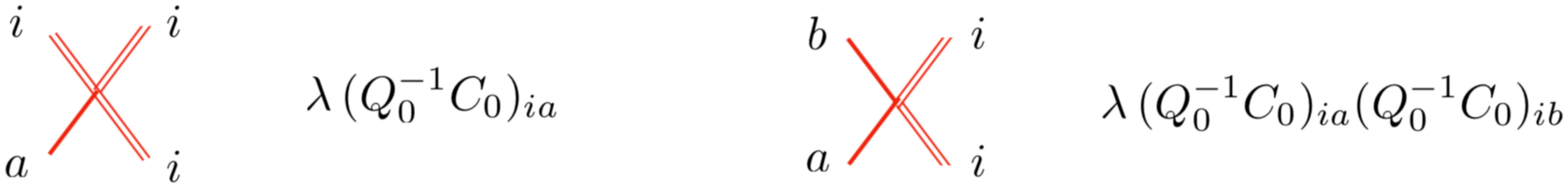}}}   \hspace{1mm} \ , \hspace{-7mm}
\label{rules4}
\end{equation}
\vspace*{-3.6cm}

\noindent  and analogously for all other possible replacements of double by single lines. In order to avoid confusion, we use the letters $a,\, b,..=1,..,8$ to label the ${\bf p}$ fields of the level one tilted lattice.
Due to the dressing $Q_0^{-1} C_0$, interaction couples ${\bf p}$ fields from different lattice links. The interaction vertex of the level one Boltzmann weight $W_1$ is
\vspace*{-2.6cm}
\begin{equation}
 \vcenter{\hbox{\hspace{-36mm} \includegraphics[width=9.2cm]{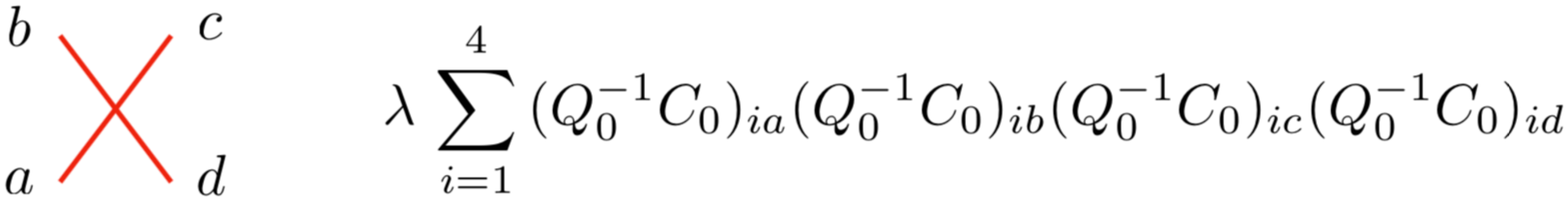}}}   \hspace{3mm} \ . \hspace{-10mm}
\label{rules5}
\end{equation}
\vspace*{-2.9cm}

The Feynman rules for constructing $F_n$ with $n>1$ are more involved due to the doubling of the splitting fields explained in Section IV.B and the different roles of the fields in the sets \eqref{chain} and \eqref{newfields}. A sketch of the Feynman rules necessary for working at first order in perturbation theory can be found in the 
readme document of the GitHub link \cite{repG}.

\section{A bound on odd fields}

Odd variables can contribute to LR mixing via the functions $F_n$, which contain the effect of interaction on the Boltzmann weights $W_n$. These functions are given by Hermite-like expressions, with the associated differential operator ${\hat {\cal F}}_n$ obtained from an integral generalizing \eqref{F2}
\begin{equation}
\hspace{-5mm} {\hat {\cal F}}_n({\bar {\bf x}},\partial_{\bf u} )\;\; = \int \! d {\bf p} \!
\vcenter{ \hbox{
\includegraphics[width=5.cm]{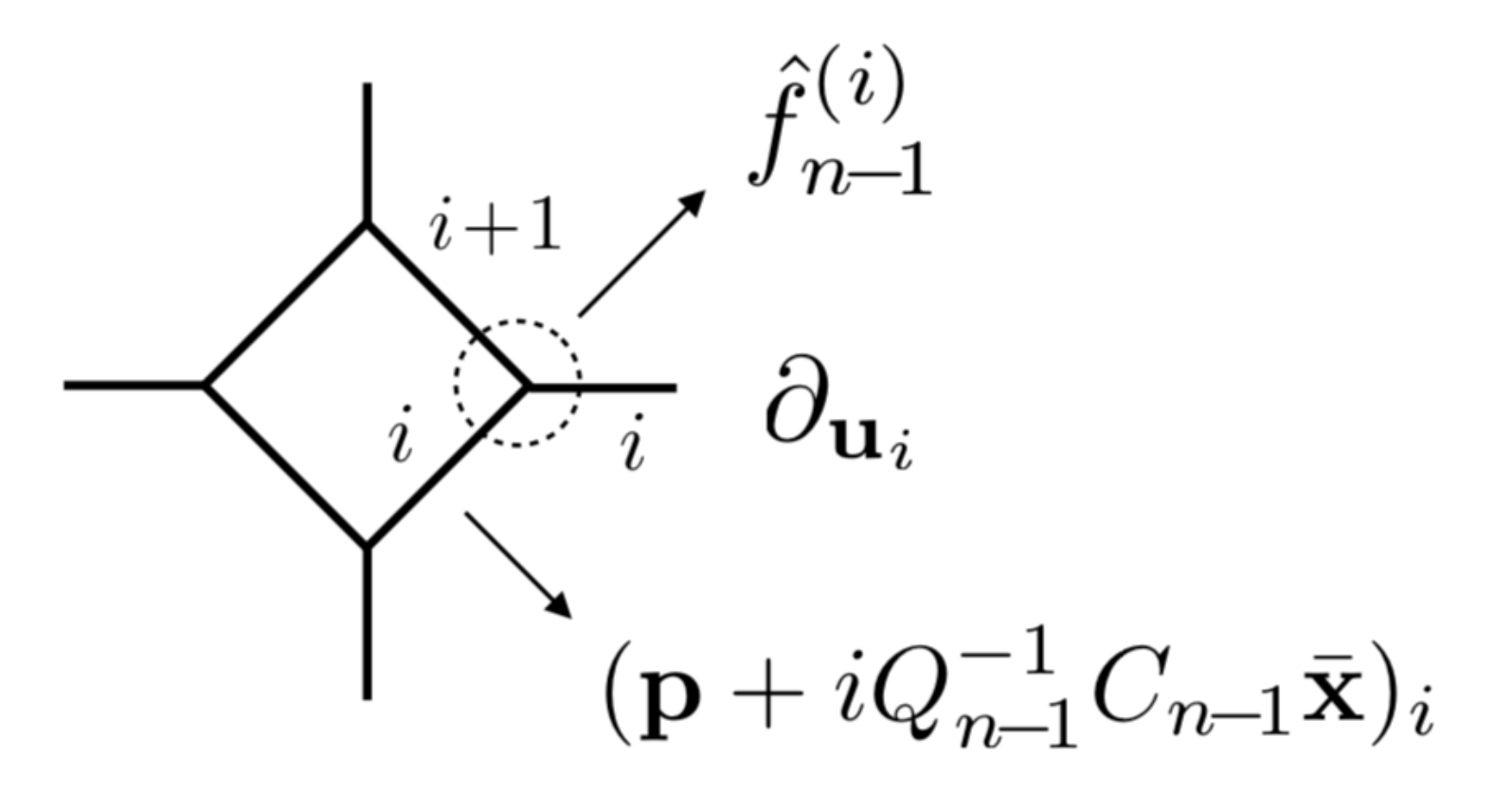}}} 
\label{Fgraph} 
\ .
\end{equation}

\noindent
The arguments of function ${\hat f}^{(i)}_{n-1}$ live at the links intersected by the dotted circle: the shifted fields ${\bf p}_{i,i+1}$ and the derivatives $\partial_{{\bf u}_i}$.
The latter generate a dependence of $F_n$ on ${\hat {\bf x}}_i$ after evaluating the corresponding expression \eqref{hermite}.

We assume that $W_n$ will be subsequently factorized along the axis $i: (1,2)_L$-$(3,4)_R$.
A generic connected Feynman diagram in the expansion of ${\hat {\cal F}}_n$ has the form

\begin{equation}
\vcenter{\hbox{
\hspace{-14mm} 
\includegraphics[width=6.2cm]{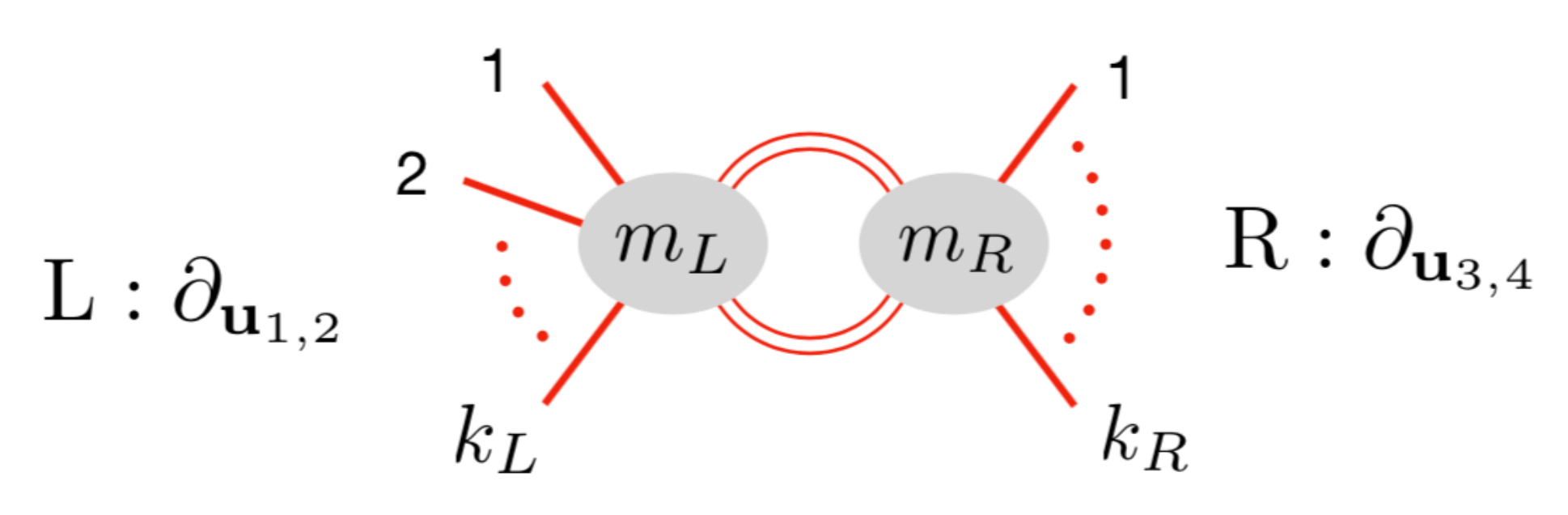}}} 
\ .
\hspace{-10mm} 
\label{LRgraph}
\end{equation}

\noindent 
The subgraph on the left combines diagrams from ${\hat f}_{n-1}^{(i)}$ with $i=1,2$ and contributes $m_L$ powers of the coupling constant. It has $k_L$ single line external legs, 
which can carry any field ${\bar {\bf x}}_j$ but only ${\bf u}$-derivatives with indices $i=1,2$. 
The same applies to the subgraph on the right, with $i=3,4$, $m_R$ and $k_R$. 
When $m_L$ or $m_R$ vanish all derivatives belong to the same L or R side.
A symmetric definition of the cubic weights emerging from $W_n$, namely $V_{n,L}=V_{n,R}$ as in \eqref{factLR}, is compatible with assigning
a diagram with $m_R$ or $m_L=0$ to $V_{n,L}$ or $V_{n,R}$ respectively. 
With this choice, factorization is achieved by just introducing splitting fields for even variables.
Hence, odd fields only require their own set of splitting fields in diagrams satisfying
\be
m_L, m_R \geq 1 \ , \;\;\;\;\;\; m_L+m_R \leq N \ ,
\label{oddfact}
\ee
with $N$ the order at which we want the perturbative expansion to stop.

\section{A bound on even fields}

Without truncation, the bond dimension in the free network doubles when progressing from the coarse graining level $2n$ to the level $2n+1$, but remains the same from $2n+1$ to $2n+2$. Namely
\be
\chi_{2n+1}=2 \chi_{2n} \ , \hspace{1cm}  \chi_{2n+2}=\chi_{2n+1} \ .
\label{chi3}
\ee
The reason behind this behaviour is the vanishing of half of the $B_{2n+1}$ singular values \cite{CS19}, which leads to discard automatically the associated splitting fields.
This property is lost once $\chi_{\rm max}$ is reached and actives the truncation preventing the bond dimension to further grow.

In the presence of interaction, a field can only be discarded when its contribution to entanglement is small both at the level of
the leading gaussian structure and the Feynman graphs. 
Related to that, in the section V.C we raised the question about the fate of the fields inherited from the free network and related to vanishing singular values of $B_{2n+1}$.
We will now prove that they do not participate in the LR factorization of graphs with $m_L$ or $m_R$ equal to zero in the notation of \eqref{LRgraph}. 

Since we are interested in odd coarse graining levels, the corresponding lattice fields will be of type ${\bf p}$. 
Analogously to \eqref{Fgraph}, the differential operator ${\hat {\cal F}}_{2n+1}$ is schematically given by 
\vspace*{-3.2cm}
\begin{equation}
\hspace{-5mm} {\hat {\cal F}}_{2n+1}({\bar {\bf p}},\partial_{\bf u} )\;\; = \int \! d {\bf x} \!
\vcenter{ \hbox{\hspace{-9mm} \includegraphics[width=6cm]{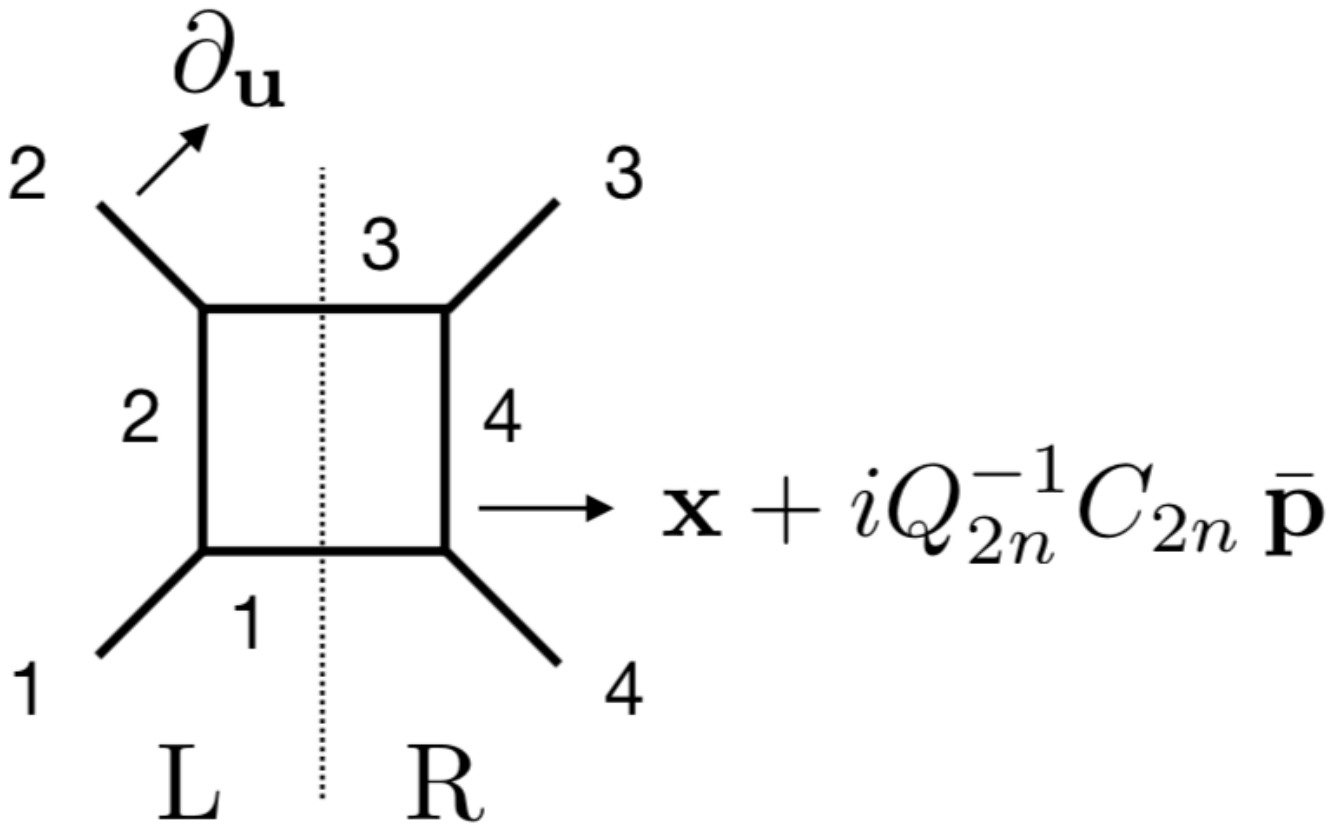}}}  \hspace{-6mm} \ . \hspace{-10mm} 
\label{Fgraphodd}
\end{equation}
\vspace*{-3.2cm}

\noindent For definiteness, we consider a Feynman graph in the expansion of  ${\hat {\cal F}}_{2n+1}$ with $m_R=0$. Such diagram only traces back to the two cubic weights on the left side of the integrand.  
Its external legs carry the expressions
\vspace*{-2.4cm}
\begin{equation}
\hspace{-4mm} \vcenter{ \hbox{\hspace{-9mm} \includegraphics[width=9.0cm]{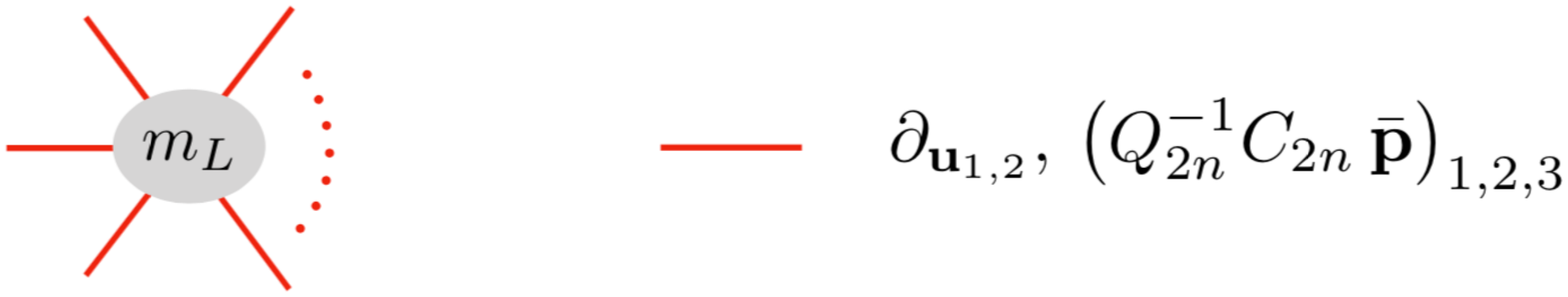}}} \hspace{-7mm}  \ .
\label{graphodd}
\end{equation}
\vspace*{-2.8cm}

\noindent 
The subindices $1,2$ of the ${\bf u}$-derivatives refer to the tilted lattice links in \eqref{Fgraphodd}, while the subindices $1,2,3$ in the linear combinations of the ${\bar {\bf p}}$ fields allude to the corresponding internal links. Contrary to the ${\bf u}$-derivatives, these linear combinations involve fields from the L and R sides and are the only source of LR cross terms in \eqref{graphodd}. 
The leading gaussian structure does not mix the fields which have a counterpart in the free network, the set \eqref{chain}, with the fields \eqref{newfields} required to keep the TRG manipulations local. Therefore although both appear in the previous linear combinations, we can treat them separately. Using this, in the following we focus on the former and disregard the latter.

It is natural to assign the graph \eqref{graphodd} to the cubic vertex $V_{2n+1,L}$. The dependence on ${\bar {\bf p}}_R$ should be thus eliminated as explained in Section IV.B. It results in
the substitution
\be
Q^{-1}_{2n} C_{2n} \,{\bar {\bf p}} = Q^{-1}_{2n} \big( C_{2n,L}\, {\bar {\bf p}}_L + C_{2n,R}\, {\bar {\bf p}}_R \big)\;  \longrightarrow \; 
Q^{-1}_{2n} \big[ (C_{2n,L}+C_{2n,R}) \, {\bar {\bf p}}_L +i \,C_{2n,R} \, U_{2n+1} \, \partial_{{\bf x}^1} \big] \ .
\label{QC}
\ee
All the matrices here are given by the expressions in Appendix B. 
In particular $C_{2n,L}$ and $C_{2n,R}$ are constructed in terms the matrices $u_{2n}$ and $v_{2n}$ obtained from the SVD of $B_{2n}$. Relations \eqref{chi3} imply that 
in the absence of truncation, $B_{2n}$ has maximal rank and thus $u_{2n}$ and $v_{2n}$ are orthogonal matrices instead of just isometries. In that case we have
\be
C_{2n,L} \; C_{2n,L}^T =  
\begin{pmatrix}
1 & 0 & 0 & 0 \\
0 & 2 & 0 & 0 \\
0 & 0 & 1 & 0 \\
0 & 0 & 0 & 0 
\end{pmatrix} 
\otimes \id_{\chi_{2n}} \ .
\ee 
The $4 \times 4$ matrix in the {\it rhs} refers to the four internal links in the integrand of \eqref{Fgraphodd}. It has rank three because $C_{2n,L}$ does not couple the fields on the left external links with those on the fourth internal link. Let ${\bf v}$ be a singular vector of $B_{2n+1}$ with zero singular value. Multiplying \eqref{CQC} by $C_{2n,L}$, we obtain
\be
\big( Q_{2n}^{-1} \, C_{2n,R} \, {\bf v} \big)_{1,2,3} =0  \ .
\ee
The rows of the matrix $U_{2n+1}$ on the {\it rhs} of \eqref{QC} contain the singular vectors of $B_{2n+1}$. Let $y^{1}$ denote the component of ${\bf x}^1$ related to ${\bf v}$. The previous equality implies that $y^{1}$ decouples from the contribution of the graph \eqref{graphodd} to the cubic vertex $V_{2n+1,L}$. 

At first order in perturbation theory $m_L +m_R =1$, and thus $m_L$ or $m_R$ must be zero. We chose again to assign the associated Feynman diagram to the L or R cubic weights respectively.
The splitting fields related to vanishing singular values of $B_{2n+1}$ 
can then be safely ignored. Therefore at first order in $\lambda$, the fields \eqref{chain} and their leading order gaussian structure coincides precisely with those of the free network.

Let us finally consider  a generic Feynman diagram in the expansion of ${\hat {\cal F}}_{2n+1}$. Its external legs can carry
\vspace*{-3.4cm}
\begin{equation}
\hspace{-0mm} \vcenter{ \hbox{\hspace{-9mm} \includegraphics[width=12.0cm]{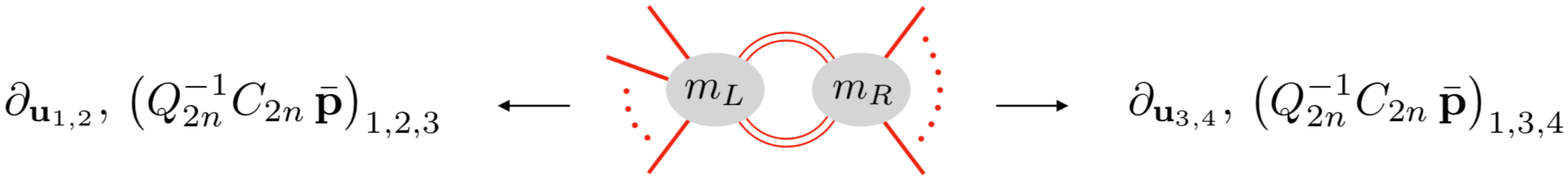}}} \hspace{-2mm}  \ .
\label{graphgen}
\end{equation}
\vspace*{-3.5cm}

\noindent If this diagram contributes to $V_{2n+1,L}$,  
the following term will appear in some of its external legs
\be
\big[( Q_{2n}^{-1} \, C_{2n,R} \, {\bf v} )_4\big] \; \partial_{y^1} \ .
\ee
The zero singular value of ${\bf v}$ does not impose the vanishing of the combination in brackets. Hence, although $y^1$ is irrelevant for \eqref{graphodd}, it can be important to transmit the entanglement created by the general graph \eqref{graphgen}. In that case the field $y^1$ should not be truncated. Analogous reasoning holds for $y^2$, the component of ${\bf x}^2$ related to ${\bf v}$, when a general graph contributes to $V_{2+1,R}$. The even combination ${\bar y}=y^1+y^2$ will enlarge then the set of fields \eqref{chain} beyond those present in the free model.


\begin{thebibliography}{99}


\bibitem{CS19}
M.~Campos, G.~Sierra and E.~Lopez,
\href{https://doi.org/10.1103/PhysRevB.100.195106}{
``Tensor renormalization group in bosonic field theory,''
Phys. Rev. B \textbf{100}, 195106  (2019).
}

\bibitem{A88}  I. Affleck, T. Kennedy, E. H. Lieb, and H. Tasaki,
\href{https://doi.org/10.1007/BF01218021}{
``Valence bond ground states in isotropic quantum antiferromagnets'',
Commun. Math. Phys., {\bf 115}, 477 (1988).
}

\bibitem{W92} S. R. White,
\href{https://doi.org/10.1103/PhysRevLett.69.2863}{
``Density matrix formulation for quantum renormalization groups'',
Phys. Rev. Lett. {\bf 69}, 2863 (1992).
}

\bibitem{F92}  M. Fannes, B. Nachtergaele, and R. F. Werner,
\href{https://doi.org/10.1007/BF02099178}{
``Finitely correlated states on quantum spin chains'',
Commun. Math. Phys. {\bf 144}, 443 (1992).
}

\bibitem{K93}  A. Kl\"{u}mper, A. Schadschneider, and J. Zittartz
\href{https://doi.org/10.1209/0295-5075/24/4/010}{
``Matrix-product-groundstates for one-dimensional spin-1 quantum antiferromagnets'',
Europhys. Lett. 24, 293 (1993).
}

\bibitem{O95} S. \"Ostlund and S.  Rommer,
\href{https://doi.org/10.1103/PhysRevLett.75.3537}{
``Thermodynamic Limit of Density Matrix Renormalization'',
Phys. Rev. Lett. {\bf 75}, 3537 (1995).
}

\bibitem{N95} T. Nishino.
\href{https://doi.org/10.1143/JPSJ.64.3598}{
``Density Matrix Renormalization Group Method for 2D Classical Models''.
J. Phys. Soc. Jpn., {\bf 64}, 3598 (1995).
}

\bibitem{NO97}
T. Nishino and K. Okunishi,
\href{https://doi.org/10.1143/JPSJ.66.3040}{
``Corner Transfer Matrix Algorithm for Classical Renormalization Group'',
J. Phys. Soc. Jpn. {\bf 66}, 3040 (1997).
}

\bibitem{D97}  J. Dukelsky, M.A. Martin-Delgado, T. Nishino, G. Sierra,
\href{https://doi.org/10.1209/epl/i1998-00381-x}{
``Equivalence of the Variational Matrix Product Method and the Density Matrix Renormalization Group applied to Spin Chains'',
Europhys. Lett., {\bf 43}, 457 (1998).
}

\bibitem{S98} G. Sierra and M.A. Martin-Delgado
\href{https://arxiv.org/abs/cond-mat/9811170}{
``The Density Matrix Renormalization Group, Quantum Groups and Conformal Field Theory'',
Proceed. Workshop on the Exact Renormalization Group, Faro (Portugal) 1998,
 arXiv:cond-mat/9811170.
}

\bibitem{V03} G. Vidal,
\href{https://doi.org/10.1103/PhysRevLett.91.147902}{
``Efficient Classical Simulation of Slightly Entangled Quantum Computations'',
Phys. Rev. Lett. {\bf 91}, 147902 (2003).
}

\bibitem{V04} F. Verstraete, D. Porras, and J. I. Cirac,
\href{https://doi.org/10.1103/PhysRevLett.93.227205}{
``Density Matrix Renormalization Group and Periodic Boundary Conditions: A Quantum Information Perspective'',
Phys. Rev. Lett. {\bf 93}, 227205 (2004).
}

\bibitem{VC04}
F. Verstraete and J. I. Cirac,
\href{https://arxiv.org/abs/cond-mat/0407066}{
``Renormalization algorithms for Quantum-Many Body Systems in two and higher dimensions'',
arXiv:cond-mat/0407066v1 (2004).
}

\bibitem{S05}  U. Schollw\"{o}ck,
\href{https://doi.org/10.1103/RevModPhys.77.259}{
``The density-matrix renormalization group'',
Rev. Mod. Phys. {\bf 77}, 259 (2005).
}

\bibitem{M05}  V. Murg, F. Verstraete, and J. I. Cirac.
\href{https://doi.org/10.1103/PhysRevLett.95.057206}{
``Efficient evaluation of partition functions of frustrated and inhomogeneous spin systems''.
Phys. Rev. Lett., {\bf 95}, 057206  (2005).
}

\bibitem{D07}  D.  P\'erez-Garc\'{\i}a, F.  Verstraete, M. M.  Wolf, J. I.  Cirac,
\href{https://dl.acm.org/doi/10.5555/2011832.2011833}{
``Matrix product state representations'',
Quantum Inf. Comput. {\bf 7}, 401 (2007).
}

\bibitem{LN07} M. Levin and C. P. Nave,
\href{https://doi.org/10.1103/PhysRevLett.99.120601}{
``Tensor Renormalization Group Approach to Two-Dimensional Classical Lattice Models'',
Phys. Rev. Lett. {\bf 99}, 120601 (2007).
}

\bibitem{V07} G. Vidal,
\href{https://doi.org/10.1103/PhysRevLett.99.220405}{
``Entanglement Renormalization'',
Phys. Rev. Lett. {\bf 99}, 220405 (2007).
}

\bibitem{G08}  V.  Giovannetti, S.  Montangero, R.  Fazio,
\href{https://doi.org/10.1103/PhysRevLett.101.180503}{
``Quantum MERA Channels'',
Phys. Rev. Lett. {\bf 101}, 180503 (2008).
}

\bibitem{V08} F. Verstraete, J.I. Cirac, V. Murg,
\href{https://doi.org/10.1080/14789940801912366}{
``Matrix Product States, Projected Entangled Pair States, and variational renormalization group methods for quantum spin systems'',
Adv. Phys. {\bf 57},143 (2008). 
}

\bibitem{P09} R. N. C. Pfeifer, G. Evenbly, and G. Vidal,
\href{https://doi.org/10.1103/PhysRevA.79.040301}{
``Entanglement renormalization, scale invariance, and quantum criticality'',
Phys. Rev. A {\bf 79}, 040301 (2009).
}

\bibitem{GW09}
Z.-C. Gu and X.-G. Wen,
\href{https://doi.org/10.1103/PhysRevB.80.155131}{
``Tensor-entanglement-filtering renormalization approach and symmetry-protected topological order'',
Phys. Rev. B {\bf 80}, 155131 (2009).
}

\bibitem{P10}  F. Pollmann, A. M. Turner, E.  Berg, and M.  Oshikawa,
\href{https://doi.org/10.1103/PhysRevB.81.064439}{
``Entanglement spectrum of a topological phase in one dimension'',
Phys. Rev. B {\bf 81}, 064439 (2010).
}

\bibitem{C11}  X.  Chen, Z.-C. Gu, and X.-G. Wen,
\href{https://doi.org/10.1103/PhysRevB.83.035107}{
``Classification of gapped symmetric phases in one-dimensional spin systems'',
Phys. Rev. B {\bf 83}, 035107 (2011).
}

\bibitem{N11}  N. Schuch, D.  P\'erez-Garc\'{\i}a, and J. I. Cirac,
\href{https://doi.org/10.1103/PhysRevB.84.165139}{
``Classifying quantum phases using matrix product states and projected entangled pair states'',
Phys. Rev. B {\bf 84}, 165139 (2011).
}

\bibitem{Sh12} Y. Shimizu,
\href{https://doi.org/10.1142/S0217732312500356}{
``Tensor renormalization group approach  to a lattice boson model'',
Mod.  Phys.  Lett.  A {\bf 27},   1250035  (2012).
}

\bibitem{O14} R. Or\'us,
\href{https://doi.org/10.1016/j.aop.2014.06.013}{
``A practical introduction to tensor networks: Matrix product states and projected entangled pair states'',
Ann. Phys. {\bf 349}, 117 (2014).
}

\bibitem{EV15}
G.  Evenbly and G.  Vidal,
\href{https://doi.org/10.1103/PhysRevLett.115.180405}{
``Tensor Network Renormalization'',
Phys. Rev. Lett. {\bf 115}, 180405 (2015). 
}

\bibitem{EV15b}
G.  Evenbly and G.  Vidal,
\href{https://doi.org/10.1103/PhysRevLett.115.200401}{
``Tensor network renormalization yields the multi-scale entanglement renormalization ansatz'',
Phys. Rev. Lett. {\bf 115}, 200401 (2015).
}

\bibitem{R17} S.-J. Ran, C. Peng, W. Li, M. Lewenstein, G. Su,
\href{https://doi.org/10.1103/PhysRevB.95.155114}{
``Criticality in Two-Dimensional Quantum Systems: Tensor Network Approach'',
 Phys. Rev. B 95, 155114 (2017).
}

\bibitem{B17} M. Bal, M. Mari\"en, J. Haegeman, F. Verstraete,
\href{https://doi.org/10.1103/PhysRevLett.118.250602}{
``Renormalization group flows of Hamiltonians using tensor networks''
Phys. Rev. Lett. 118, 250602 (2017).
}

\bibitem{H18} H.  He, Y. Zheng, B. Andrei Bernevig, N.  Regnault,
\href{https://doi.org/10.1103/PhysRevB.97.125102}{
``Entanglement Entropy From Tensor Network States for Stabilizer Codes'',
Phys. Rev. B 97, 125102 (2018).
}

\bibitem{S19} S. Singha Roy, H. Shekhar Dhar, A. Sen De, U. Sen,
\href{https://doi.org/10.1103/PhysRevA.99.062305}{
``Tensor-network approach to compute genuine multisite entanglement in infinite quantum spin chains'',
Phys. Rev. A 99, 062305 (2019).
}

\bibitem{C19} M. C.  Banuls, K.  Cichy, H.-T.  Hung, Y.-J.  Kao, C.-J. D. Lin, Y.-P. Lin, D. T.-L. Tan,
\href{https://doi.org/10.22323/1.363.0022}{
``Phase structure and real-time dynamics of the massive Thirring model in 1+1 dimensions using the tensor-network method'',
 PoS (LATTICE2019) 022.
}

\bibitem{K19}
D.~Kadoh, Y.~Kuramashi, Y.~Nakamura, R.~Sakai, S.~Takeda and Y.~Yoshimura,
\href{https://doi.org/10.1007/JHEP05(2019)184}{
``Tensor network analysis of critical coupling in two dimensional $\phi^{4}$ theory'',
JHEP \textbf{05} (2019), 184.
}
 
\bibitem{V19} 
B. Vanhecke, J.  Haegeman, K. Van Acoleyen, L. Vanderstraeten, F. Verstraete, 
\href{https://doi.org/10.1103/PhysRevLett.123.250604}{
``A scaling hypothesis for matrix product states'', 
 Phys. Rev. Lett. {\bf 123}, 250604 (2019). 
}

\bibitem{G19} J.  Garre-Rubio, 
\href{https://arxiv.org/abs/1912.08597}{
``Symmetries in topological tensor network states: classification, construction and detection'', arXiv:1912.08597.
}

\bibitem{B20} M. C. Banuls, M. P. Heller, K. Jansen, J. Knaute, V. Svensson,
\href{https://doi.org/10.1103/PhysRevResearch.2.033301}{
``From spin chains to real-time thermal field theory using tensor networks'',
Phys. Rev. Research {\bf 2}, 033301 (2020).
}

\bibitem{J20} Hui-Ke Jin, Hong-Hao Tu, Yi Zhou,
\href{https://doi.org/10.1103/PhysRevB.101.165135}{
``Efficient tensor network representation for Gutzwiller projected states of paired fermions'',
Phys. Rev. B {\bf 101}, 165135 (2020)
}

\bibitem{D20} C. Delcamp, A. Tilloy, 
\href{https://doi.org/10.1103/PhysRevResearch.2.033278}{
``Computing the renormalization group flow of two-dimensional $\phi^4$ theory with tensor networks'',
Phys. Rev. Research 2, 033278 (2020);
}

\bibitem{M20} Q. Mortier, N.Schuch, F. Verstraete, J. Haegeman,
\href{https://arxiv.org/abs/2008.11176}{
``Resolving Fermi surfaces with tensor networks'', 
arXiv:2008.11176.
}

\bibitem{P20} D. Poilblanc, M. Mambrini, F. Alet,
\href{https://scipost.org/10.21468/SciPostPhys.10.1.019}{
``Finite-temperature symmetric tensor network for spin-1/2 Heisenberg antiferromagnets on the square lattice'',
SciPost Phys. {\bf 10}, 019 (2021).
}

\bibitem{V21} B. Vanhecke, F. Verstraete, K. Van Acoleyen, 
\href{https://arxiv.org/abs/2104.10564}{
``Entanglement scaling for $\lambda \phi_2^4$'', 
arxiv.2104.10564.
}




\bibitem{V10}  F. Verstraete and J. I. Cirac,
\href{https://doi.org/10.1103/PhysRevLett.104.190405}{
``Continuous matrix product states for quantum fields'',
Phys. Rev. Lett. {\bf 104}, 190405 (2010).
}

\bibitem{H13}  J.  Haegeman, T. J. Osborne, H. Verschelde and F. Verstraete,
\href{https://doi.org/10.1103/PhysRevLett.110.100402}{
``Entanglement renormalization for quantum fields in real space'',
Phys. Rev. Lett. {\bf 110}, 100402 (2013).
}

\bibitem{J15}  D. Jennings, C. Brockt, J. Haegeman, T. J.  Osborne and F.  Verstraete,
\href{https://doi.org/10.1088/1367-2630/17/6/063039}{
``Continuum tensor network field states, path integral representations and spatial symmetries'',
New J. Phys. {\bf 17}, 063039 (2015).
}

\bibitem{TC18}
A. Tilloy, J. I.  Cirac
\href{https://doi.org/10.1103/PhysRevX.9.021040}{
``Continuous Tensor Network States for Quantum Fields'',
Phys. Rev. X {\bf 9}, 021040 (2019)
}

\bibitem{CM18}  J. Cotlera, M. R. M. Mozaffar, A. Mollabashi, A. Naseh,
\href{https://doi.org/10.1002/prop.201900038}{
``Renormalization Group Circuits for Weakly Interacting Continuum Field Theories'',
Fortschr. Phys. {\bf 67}, 1900038 (2019).
}

\bibitem{HV18}  Q.  Hu, A. Franco-Rubio, G. Vidal,
\href{https://arxiv.org/abs/1809.05176}{
``Continuous tensor network renormalization for quantum fields'',
arXiv:1809.05176.
}

\bibitem{K20} T.  D. Karanikolaou, P.  Emonts, A.  Tilloy, 
\href{https://doi.org/10.1103/PhysRevResearch.3.023059}{
``Gaussian Continuous Tensor Network States for Simple Bosonic Field Theories''
Phys. Rev. Research {\bf 3}, 023059 (2021). 
arXiv:2006.13143.
}

\bibitem{N21} A.E. B. Nielsen, B. Herwerth, J. I. Cirac, and G. Sierra,
\href{https://doi.org/10.1103/PhysRevB.103.155130}{
``Field tensor network states'',
Phys. Rev. B 103, 155130 (2021).
}

\bibitem{T21a}
A. Tilloy,
\href{https://doi.org/10.1103/PhysRevD.104.L091904}{
``Variational method in relativistic quantum field theory without cutoff'',
Phys. Rev. D 104, L091904 (2021).
}

\bibitem{T21b}
A. Tilloy,
\href{https://arxiv.org/abs/2102.07741}{
``Relativistic continuous matrix product states for quantum fields without cutoff'',
arXiv:2102.07741.
}




\bibitem{Sw12}  B. Swingle,
\href{https://doi.org/10.1103/PhysRevD.86.065007}{
``Entanglement renormalization and holography'',
Phys. Rev. D {\bf 86}, 065007 (2012).
}

\bibitem{LS15} J. I. Latorre and G. Sierra, 
\href{https://arxiv.org/abs/1502.06618}{
``Holographic codes'', arXiv:1502.06618.
}

\bibitem{PY15}  F.  Pastawski, B.  Yoshida, D.  Harlow, and John Preskill,
\href{https://doi.org/10.1007/JHEP06(2015)149}{
``Holographic quantum error-correcting codes: toy models for the bulk/boundary correspondence'',
J. High Energy Phys. 2015,  {\bf 149},   (2015).
}

\bibitem{Mo15}  J. Molina-Vilaplana,
\href{https://doi.org/10.1007/JHEP09(2015)002}{
``Information geometry of entanglement renormalization for free quantum fields'',
J. High Energ. Phys. 2015, 2 (2015).
}

\bibitem{MN15}  M. Miyaji, T. Numasawa, N. Shiba, T. Takayanagi, K. Watanabe,
\href{https://doi.org/10.1103/PhysRevLett.115.171602}{
``cMERA as Surface/State Correspondence in AdS/CFT'',
Phys. Rev. Lett. {\bf 115}, 171602 (2015). 
}

\bibitem{CK17}  P.  Caputa, N.  Kundu, M.  Miyaji, T.  Takayanagi, and K.  Watanabe,
\href{https://doi.org/10.1007/JHEP11(2017)097}{
``Liouville action as path-integral complexity: from continuous tensor networks to AdS/CFT'',
J. High Energy Phys. 2017, {\bf 97}  (2017).
}

\bibitem{VP19} R.  Vasseur, A.  C. Potter, Y.-Z.  You, A. W. W. Ludwig,
\href{https://doi.org/10.1103/PhysRevB.100.134203}{
``Entanglement Transitions from Holographic Random Tensor Networks'',
 Phys. Rev. B {\bf 100}, 134203 (2019). 
}

\bibitem{J21} A.  Jahn, J. Eisert, 
\href{https://doi.org/10.1088/2058-9565/ac0293}{
``Holographic tensor network models and quantum error correction: A topical review'',
Quantum Sci. Technol. 6 033002 (2021).
}


\bibitem{repG} 
\href{https://github.com/m-campos/interacting-trg}{https://github.com/m-campos/interacting-trg}




\bibitem{W75}  K.G. Wilson, 
\href{https://doi.org/10.1103/PhysRevB.4.3174}{
``Group and Critical Phenomena. I. Renormalization Group and the Kadanoff Scaling Picture'',
Phys. Rev. B {\bf 4}, 3174 (1971).
}

\bibitem{S94} R. Shankar, 
\href{https://doi.org/10.1103/RevModPhys.66.129}{
``Renormalization Group Approach to Interacting Fermions'',
Rev. Mod. Phys. {\bf 66}, 129 (1994).
}

\bibitem{A70}
M. Abramowitz, I. A.  Stegun, 
\href{https://dl.acm.org/doi/10.5555/1098650}{
``Handbook of Mathematical Functions with Formulas, Graphs, and Mathematical Tables'',
Applied Mathematics Series {\bf 55} , New York,  Dover Publications (1970).
}




\end{thebibliography}
\end{document}